# Torsional parallel plate flow of Herschel-Bulkley fluids with wall slip

Evgenios Gryparis
*Department of Mathematics and Statistics,*
*University of Cyprus, PO Box 20537, 1678, Nicosia, Cyprus*
E-mail: gryparis.evgenios@ucy.ac.cy

Georgios C. Georgiou*
*Department of Mathematics and Statistics,*
*University of Cyprus, PO Box 20537, 1678, Nicosia, Cyprus*
E-mail: georgios@ucy.ac.cy
*: Corresponding author

## Abstract

The effect of wall slip on the apparent flow curves of viscoplastic materials obtained using torsional parallel plate rheometers is analysed by considering Herschel-Bulkley fluids and assuming that slip occurs above a critical wall shear stress, the slip yield stress $\tau_c$, taken to be lower than the yield stress, $\tau_0$. Thus, different flow regimes are encountered as the angular velocity of the experiment is increased. When the rim shear stress $\tau_R$ is below the slip yield stress, the exerted torque is not sufficient to rotate the disk and the material remains still. When $\tau_c < \tau_R \le \tau_0$ the material is still unyielded but exhibits wall slip and rotates as a solid at half the angular velocity of the rotating disk. Finally, when $\tau_R > \tau_0$, the material exhibits slip everywhere and yields only in the annulus $r_0 \le r \le R$, where $r_0$ is the critical radius at which the shear stress is equal to the yield stress and $R$ is the radius of the disks. In the general case, the slip velocity, which varies with the radial distance, can be calculated numerically and then all quantities of interest, such as the true shear rate, and the two branches of the apparent flow curve can be computed by means of closed form expressions. Analytical solutions have also been obtained for certain values of the power-law exponent. In order to illustrate the effect of wall slip on the apparent flow curve and on the torque, results have been obtained for different gap sizes between the disks choosing the values of the rheological and slip parameters to be similar to reported values for certain colloidal suspensions. The computed apparent flow curves reproduce the patterns observed in the experiments.

## 1. Introduction

Wall slip has attracted considerable attention in the past few decades since it affects dramatically the stability of several flows of industrial interest and the accuracy of rheological measurements. The reader is referred to the recent reviews of Hatzikiriakos (2015), Cloitre and Bonnecaze (2017), and Malkin and Patlazhan (2018). Various slip laws have been proposed in the literature, which replace the classical no-slip condition that dictates that fluid particles stick at the wall, i.e., they move with the same speed as the wall. These laws relate the slip velocity $u_w$, defined as the relative velocity of the fluid particles with respect of that of the wall, to the wall shear stress, $\tau_w$. The most common slip equation is that proposed by Navier (1827):

$$\tau_w = \beta u_w \tag{1}$$

where $\beta$ is the slip coefficient. In general, $\beta$ depends on the fluid and wall properties, the temperature, the normal stress, and the pressure (Hatzikiriakos, 2015). Equation (1) tends asymptotically to the no-slip boundary condition when $\beta \to \infty$ ($u_w = 0$). Its power-law generalization,

$$\tau_w = \beta u_w^m \tag{2}$$

where $m$ is the slip exponent, is also widely used (Hatzikiriakos, 2015), e.g., for highly concentrated suspensions (Wilms et al., 2022) and viscoplastic microgels (Medina-Bañuelos et al., 2017).

Most experimental studies have demonstrated that wall slip occurs only above critical value of the wall shear stress, known as the slip yield stress, $\tau_c$ (Malkin and Patlazhan, 2018; for additional references see Damianou et al., 2019). It should be noted that non-zero slip-yield-stress values have been reported not only for complex but also for Newtonian fluids (Spikes and Granick, 2003). The above slip equations have been extended to account for the slip yield stress. For example, Eq. (2) is generalized as follows:

$$\begin{cases} u_w = 0, & \tau_w \le \tau_c \\ \tau_w = \tau_c + \beta u_w^m, & \tau_w > \tau_c \end{cases} \tag{3}$$

More complex slip equations involving finite slip yield stress have also been proposed; see, for example, the non-monotonic slip equations proposed by Piau and El Kissi (1994) for certain polymer melts and the review in Damianou et al. (2014).

The two-branch nature of slip laws with non-zero slip yield stress results in different flow regimes in viscometric flows. For example, in simple shear (plane Couette) flow there is critical speed of the moving plate separating the no-slip from the slip regime (Georgiou, 2021). Additional flow regimes arise in one-dimensional flows, characterized by two characteristic wall shear rates, e.g., in circular (Damianou et al., 2019) and annular (Gryparis and Georgiou, 2022) Couette flows, or in two-dimensional flows, e.g., in pressure-driven flow in a rectangular duct (Damianou and Georgiou, 2014)

Of special interest are viscoplastic flows, i.e., flows of materials with a finite yield stress, $\tau_y$. These include many classes of materials of industrial importance, such as pastes, cements, mortars, foams, muds, food products, etc. (Coussot et al., 2014). Ideal yield-stress fluids behave as solids if the stress is below the yield stress and as fluids otherwise (Coussot et al., 2014). The constitutive equation proposed by Bingham (1922) is widely used. The scalar form of this equation is as follows:

$$\begin{cases} \dot{\gamma} = 0, & \tau \le \tau_y \\ \tau = \tau_y + \mu\dot{\gamma}, & \tau > \tau_y \end{cases} \tag{4}$$

where $\tau$ is the shear stress, $\dot{\gamma}$ is the shear rate, $\tau_y$ is the yield stress, and $\mu$ is the plastic viscosity. Equation (4) reduces to the Newtonian constitutive equation when $\tau_y = 0$. It has been extended by Herschel and Bulkley (1926) to

$$\begin{cases} \dot{\gamma} = 0, & \tau \le \tau_y \\ \tau = \tau_y + k\dot{\gamma}^n, & \tau > \tau_y \end{cases} \tag{5}$$

where $k$ is the consistency index and $n$ is the power-law exponent or flow index. By setting $\tau_y = 0$, Eq. (5) reduces to the power-law model.

There is ample experimental evidence that viscoplastic materials are prone to wall slip (Cloitre and Bonnecaze, 2017). In fact, wall slip may be exhibited below the yield stress, which implies that the slip yield stress is lower than the yield stress ($\tau_c < \tau_y$); see experimental data on hard-sphere colloidal suspensions (Ballesta et al., 2012) and Carbopol gels (Piau, 2017), as well the literature review provided by Damianou et al. (2019). It is clear that different regimes are encountered in viscoplastic flows with wall slip obeying a slip law with non-zero slip yield stress. In previous works, a number of basic viscoplastic flows with wall slip and non-zero slip yield stress have been analyzed, such as Poiseuille flows in pipes (Damianou et al., 2014), rectangular ducts (Damianou and Georgiou, 2014) and annuli (Gryparis and Georgiou, 2022), and the plane (Georgiou, 2021; Huilgol and Georgiou, 2022) and circular Couette flows (Damianou et al., 2019).

The occurrence of wall slip greatly affects the apparent flow curves in all rheometric flows. Thus, the apparent flow curves are diameter-dependent in capillary rheometers and gap-dependent in circular Couette or parallel plate rheometers. Therefore, studying wall slip is of utmost importance in correcting rheometric data obtained using different rheometers and geometries in order to determine the true rheology of materials. Mooney (1931) proposed a methodology to analyse rheological data from capillary and circular Couette rheometers and derived convenient explicit formulae for the determination of the slip velocity as a function of wall shear stress. Schofield and Scott Blair (1931) derived explicit formulae to calculate the slip velocity from capillary experimental data. Subsequently, Schofield (1934) derived simple relations to address and explain discrepancies of flow curves due to wall slip. Oldroyd (1949) also proposed a similar methodology for determining the slip velocity as a function of the wall shear stress and recovering the true rheological parameters. Based on Oldroyd's ideas, Ghahramani et al. (2021) recently derived expressions for slip analysis of capillary rheometer data on Herschel-Bulkley fluids. Yoshimura and Prud'homme (1988) also analysed wall slip in Couette and parallel disc viscometers. As discussed in Section 2, in their analysis the slip velocity is assumed to vary with the wall shear stress and can be determined by carrying out experiments with two different gap sizes.

All the above analyses are general, i.e., they hold for any fluid, and have been applied extensively in analysing experimental data on polymer melts and solutions as well as on many other complex fluids (Barnes, 1995; Hatzikiriakos, 2015). With viscoplastic materials, the shape of the apparent flow curve may suddenly change at the transitions from the no-yielding to the slip regime and then to the yielding regime. The data at the transition points may then be used for determining the slip yield stress and the yield stress (Moud et al., 2022).

Moud et al. (2022) have carried out parallel-plate experiments with different gaps in order to characterize the wall slip of colloidal kaolinite suspensions, using both smooth and rough plates, which correspond to the no-slip and slip cases, respectively. They have identified two slip regimes below and above the yield stress. In the first regime, the material slips like an elastic solid and in the latter one the material yields and flows following a different slip law. The two slip laws were coupled with the Herschel-Bulkley constitutive equation, and the rheological and slip parameters have been calculated by numerically fitting all data (corresponding to different gap sizes). The numerical method allows the correct calculation of the yield stress value, confirmed with data obtained from parallel-plate, cone-and-plate, and concentric cylinder rheometers.

The flow of a Herschel-Bulkley fluid in a parallel-plate rheometer in the presence of wall slip with non-zero slip yield stress, which is lower than the yield stress, has been considered in Huilgol and Georgiou (2022) under the assumption that the same slip law applies independently of the yielding status of the material. As mentioned above, Moud et al. (2022) employed different slip laws in yielded and unyielded regions demanding continuity of the slip velocity, which leads to a constraint between the slip parameters in the two regions, including the slip yield stress. Huilgol and Georgiou (2022) formulated the general flow equations and provided analytical solutions for the special case when $n=m=1$ (Bingham plastic flow).

Quan et al. (2023) also studied the torsional flow of a viscoplastic hydrogel with wall slip with zero slip yield stress and presented both experimental and computational results. They reported two flow regimes below and above the critical torque at which the material yields and derived analytical expressions for the torque when the material is still unyielded and when the apparent shear rate is sufficiently large, so that wall slip is considered negligible. They also discussed the dependence of the slope of the torque as a function of the apparent rim shear rate on the rheological parameters and the flow conditions.

The objective of the present work is to analyse the flow in the general case of a Herschel-Bulkley fluid and to investigate the effects of the gap size and wall slip on the apparent flow curve, which is more relevant to experimental studies. An analogous study has been recently carried out in Georgiou (2021) for the simple shear (plane Couette) flow. Eventhough the basic results for the apparent flow curves, i.e., the plots of the rim shear stress versus the apparent shear rate, are, of course, equivalent, the characteristics of the two-dimensional flow field and the different flow regimes that arise in torsional flow are worthy of investigation. The present analysis differs from that of Moud et al. (2022) in that the same slip law is assumed to hold uniformly below and above the yield stress. Moreover, analytical expressions for the torque are obtained in terms of the apparent rim shear rate and the determination of the rheological and slip parameters are discussed.

The equations governing the general flow are presented in Section 2. The presence of two critical stress values, i.e., the yield stress and the slip yield stress, results in the appearance of different flow regimes as the angular velocity of the rotating disk is increased. In Section 3, the flow of a power-law fluid is considered first allowing different slip laws but with the same slip yield stress along the two disks (thus only the slip coefficients and the slip exponents may be different). When the rim shear stress $\tau_R = \tau_{z\theta}(R)$ is below the slip yield stress $\tau_c$, there is no wall slip and the standard textbook solution is obtained. Once $\tau_R$ exceeds $\tau_c$, wall slip does occur but only in the annulus $r_c \le r \le R$, where $r_c$ is the radius where $\tau_{z\theta}(r_c)=\tau_c$; the no-slip solution still holds for $0 \le r \le r_c$. In the general case, the slip velocities corresponding to a given radial distance are different and need to be calculated numerically solving a simple non-linear equation. Closed form expressions in terms of one of the two slip velocities can be obtained for the other slip velocity, the azimuthal velocity, the true shear rate, and the shear stress. However, full analytical solutions are obtained for special cases of the power-law and slip exponents, e.g., for Navier slip, or for special combinations of the slip exponents. The flow of a Herschel-Bulkley fluid is considered in Section 4. For the sake of simplicity, it is assumed that the same slip law with non-zero slip yield stress ($\tau_c \le \tau_y$) applies at the two disks. It should be noted that the upper disk starts rotating only when the rim shear stress exceeds the slip yield stress. When $\tau_c < \tau_R \le \tau_y$ the material remains unyielded, rotating at half the angular velocity of the rotating disk, due to wall slip. Once $\tau_R$ exceeds $\tau_y$, the material continues to slip everywhere and yields only in the annulus $r_0 \le r \le R$, $r_0$ being the yield radius (where $\tau_{z\theta}(r_0)=\tau_y$). Again, the slip velocity which varies with the radial distance is calculated numerically, whereas the azimuthal velocity, the true shear rate, the shear stress, and the torque can be computed by means of closed form formulas. Analytical solutions can also be obtained for special cases of the power-law and slip exponents, e.g., for Bingham plastics and/or Navier slip. The effect of wall slip on the apparent flow curve is illustrated by carrying out calculations for different gap sizes between the disks and by comparing the computed apparent flow curves with experimental observations on certain colloidal suspensions. Finally, the conclusions of this work are summarized in Section 5.

## 2. Analysis of the flow

We consider the torsional flow between parallel concentric disks employing cylindrical coordinates $(r, z)$, with the origin set at the center of the bottom disk, as shown in Fig. 1. It is assumed that the lower disk is fixed and that the upper disk rotates at an angular speed $\Omega$ around the common symmetry axis, and that the gap $H$ between the two disks is narrow so that the stress $\tau_{z\theta}$ for a given radial distance $r$ is approximately constant (Yoshimura and Prud'homme, 1988), i.e., $\tau_{z\theta} = \tau_{z\theta}(r)$. Hence, for a given value of $r$, the angular velocity $u_\theta$ varies linearly with $z$. If different slip laws apply at the two plates, the slip velocities at the two plates are different. Hence, the boundary conditions at the two disks are:

$$u_\theta(r,0) = u_{w1}(r) \quad \text{and} \quad u_\theta(r,H) = \Omega r - u_{w2}(r) \tag{6}$$

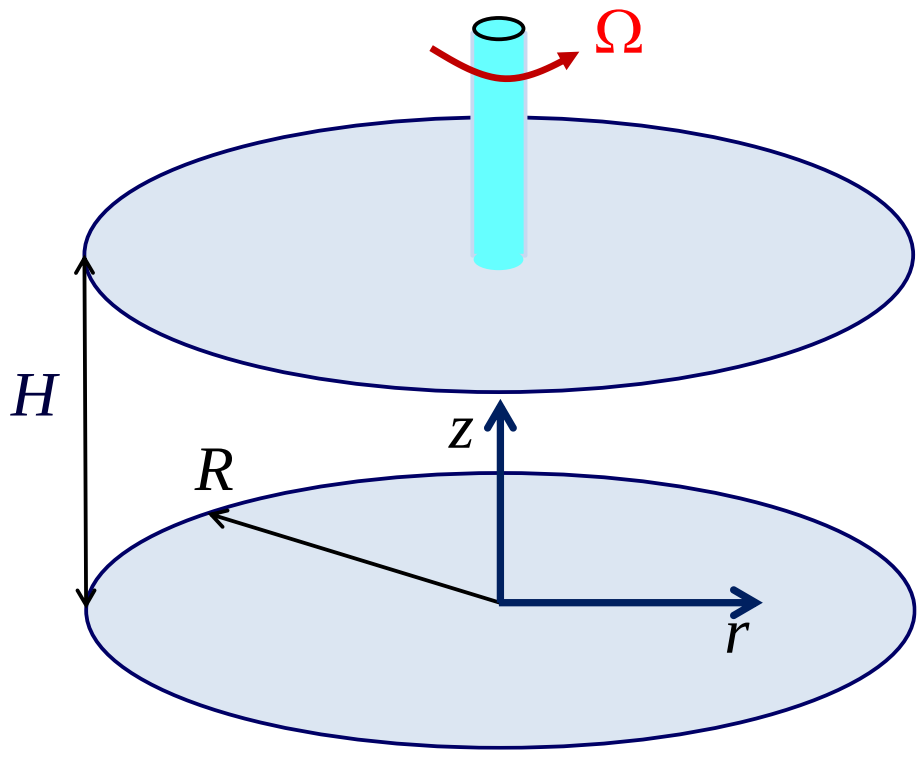

**Figure 1**. Geometry of the torsional parallel plate flow.

where $u_{w1}(r)$ and $u_{w2}(r)$ are the slip-velocity functions along the lower ($i=1$) and the upper ($i=2$) disk, respectively. It turns out that the angular velocity $u_\theta(r,z)$ is given by

$$u_\theta(r,z) = \frac{\Omega r - u_{w1}(r) - u_{w2}(r)}{H} z + u_{w1}(r) \tag{7}$$

which can also be written as follows:

$$u_\theta(r,z) = \left[ \dot{\gamma}_a(r) - \frac{u_{w1}(r) + u_{w2}(r)}{H} \right] z + u_{w1}(r) \tag{8}$$

where

$$\dot{\gamma}_a = \frac{\Omega r}{H} \tag{9}$$

is the apparent shear rate (Macosko, 1994). Thus, for the true shear rate one gets:

$$\dot{\gamma}(r) = \dot{\gamma}_a(r) - \frac{u_{w1}(r) + u_{w2}(r)}{H} \tag{10}$$

Let us note that the apparent and the true shear rates at the rim ($r = R$) are given by

$$\dot{\gamma}_{aR} = \dot{\gamma}_a(R) = \frac{\Omega R}{H} \tag{11}$$

and

$$\dot{\gamma}_R = \dot{\gamma}_{aR} - \frac{u_{w1R} + u_{w2R}}{H} \tag{12}$$

where the subscript $R$ denotes quantities at $r = R$. It is clear that in the case of no slip, the apparent shear rate coincides with the true one

$$\dot{\gamma}(r) = \dot{\gamma}_a(r) \tag{13}$$

and the azimuthal velocity is given by

$$u_\theta(r,z) = \frac{\Omega r\, z}{H} = \dot{\gamma}_a(r) z \tag{14}$$

(hence, $\dot{\gamma}_{aR}$ and $\dot{\gamma}_R$ coincide only in the absence of wall slip).

If the same slip law applies at the two plates, Eqs. (8) and (10) are simplified to (Yoshimura and Prud'homme, 1988)

$$u_\theta(r,z) = \left[\dot{\gamma}_a(r) - \frac{2u_w(r)}{H}\right] z + u_w(r) \tag{15}$$

or

$$u_\theta(r,z) = \dot{\gamma}_a(r) z + u_w(r)\left(1 - \frac{2z}{H}\right) \tag{16}$$

and

$$\dot{\gamma}(r) = \dot{\gamma}_a(r) - \frac{2u_w(r)}{H} \tag{17}$$

where $u_w(r)$ is the common slip velocity at the two disks.

The torque, $M$, required to observe torsional flow at a given apparent rim shear rate, $\dot{\gamma}_{aR}$, is given by:

$$M = 2\pi \int_0^R \tau_{z\theta}(r) r^2 dr\,, \tag{18}$$

which by means of (9) gives

$$M = \frac{2\pi R^3}{\dot{\gamma}_{aR}^3} \int_0^{\dot{\gamma}_{aR}} \tau_{z\theta}(\dot{\gamma}_a) \dot{\gamma}_a^2 d\dot{\gamma}_a\,. \tag{19}$$

Application of Leibniz's rule leads to the well-known formula for the rim shear stress $\tau_R$ (Macosko, 1994):

$$\tau_R = \frac{M}{2\pi R^3}\left(3 + \frac{d \ln M}{d \ln \dot{\gamma}_{aR}}\right) \tag{20}$$

Yoshimura and Prud'homme (1988) showed that in the presence of slip the true shear rate and thus the viscosity can be determined using data obtained for two gap heights $H_1$ and $H_2$. Setting $r = R$ in Eq. (17), one gets

$$\dot{\gamma}_R = \dot{\gamma}_{aRj} - \frac{2u_{w_R}}{H_j}, \quad j = 1,2 \tag{21}$$

and, thus, the rim slip velocity for a given shear stress at the edge of the upper disk, $\tau_R = \tau_{z\theta}(R)$, is given by

$$u_{wR}(\tau_R) = \frac{\dot{\gamma}_{aR1}(\tau_R) - \dot{\gamma}_{aR2}(\tau_R)}{2\left(\dfrac{1}{H_1} - \dfrac{1}{H_2}\right)} \tag{22}$$

The corresponding true shear rate is (Yoshimura and Prud'homme, 1988)

$$\dot{\gamma}_R(\tau_R) = \frac{H_1\dot{\gamma}_{aR1}(\tau_R) - H_2\dot{\gamma}_{aR2}(\tau_R)}{H_1 - H_2} \tag{23}$$

The above analysis is general, since it holds for any fluid and is independent of the slip equations that apply at the two walls. In the next two sections we will first consider the flow of a power-law fluid with different slip equations and then the flow of a Herschel-Bulkley fluid with the same slip equation holding at the two disks.

## 3. Torsional flow of a power-law fluid

Consider the torsional flow of a power-law fluid assuming that different slip laws apply at the two disks:

$$\left.\begin{array}{ll} u_{wi} = 0, & \tau_{wi} \le \tau_c \\ \tau_{wi} = \tau_c + \beta_i u_{wi}^{m_i}, & \tau_{wi} > \tau_c \end{array}\right\}, \qquad i = 1,2 \tag{24}$$

where the subscripts 1 and 2 denote quantities corresponding to the lower and upper disks, respectively. For the sake of simplicity, the value of the slip yield stress ($\tau_c$) has been taken to be the same at both disks. Under these assumptions, two regimes are encountered as the angular velocity of the rotating disk, or, equivalently, the rim shear stress $\tau_R = \tau_{z\theta}(R)$ is increased:

(i) When $\tau_R \le \tau_c$, no wall slip is observed and hence one obtains the standard no-slip solution, given by Eqs. (13) and (14). By means of the power-law constitutive equation, the shear stress and the rim shear stress are given by

$$\tau_{z\theta}(r) = k\dot{\gamma}_a^n(r) = k\left(\frac{\Omega r}{H}\right)^n \tag{25}$$

and

$$\tau_R = k\dot{\gamma}_{aR}^n = k\left(\frac{\Omega R}{H}\right)^n \tag{26}$$

The critical angular velocity $\Omega_c$ above which wall slip occurs corresponds to $\tau_R = \tau_c$, which gives

$$\Omega_c = \frac{H}{R}\left(\frac{\tau_c}{k}\right)^{1/n} \tag{27}$$

The corresponding critical apparent shear rate at the rim is

$$\dot{\gamma}_{ac} = \frac{\Omega_c R}{H} = \left(\frac{\tau_c}{k}\right)^{1/n} \tag{28}$$

By combining Eqs. (18) and (25) the torque in this regime is found to be

$$M = \frac{2\pi R^3 k \dot{\gamma}_{aR}^n}{n+3}, \quad \dot{\gamma}_{aR} \le \dot{\gamma}_{ac}\,. \tag{29}$$

(ii) When $\tau_R > \tau_c$, (or $\Omega > \Omega_c$ ), slip does occur but only in the annulus $r_c \le r \le R$, where $r_c$ is the critical radius at which $\tau_{z\theta}(r_c) = \tau_c$; in the core cylinder $0 \le r \le r_c$, the no- slip solution still applies, and therefore Eq. (25) yields

$$r_c = \frac{H}{\Omega}\left(\frac{\tau_c}{k}\right)^{1/n} \tag{30}$$

In summary, the azimuthal velocity and the shear stress are given in terms of the two slip velocities by

$$u_\theta(r,z) = \begin{cases} \dot{\gamma}_a(r) z, & 0 \le r \le r_c \\ \left[\dot{\gamma}_a(r) - \dfrac{u_{w1}(r) + u_{w2}(r)}{H}\right] z + u_{w1}(r), & r_c \le r \le R \end{cases} \tag{31}$$

and

$$\tau_{z\theta}(r) = \begin{cases} k\dot{\gamma}_a^n, & 0 \le r \le r_c \\ k\left[\dot{\gamma}_a(r) - \dfrac{u_{w1}(r) + u_{w2}(r)}{H}\right]^n, & r_c \le r \le R \end{cases} \tag{32}$$

Another quantity of interest is the rim shear stress for which we get:

$$\tau_R = k\left(\dot{\gamma}_{aR} - \frac{u_{w1R} + u_{w2R}}{H}\right)^n \tag{33}$$

The unknown slip velocities in the annulus $r_c \le r \le R$ are calculated by means of Eq. (32) and the slip equations (24):

$$\tau_c + \beta_i u_{wi}^{m_i} = k\left[\dot{\gamma}_a(r) - \frac{u_{w1}(r) + u_{w2}(r)}{H}\right]^n, \qquad i = 1,2 \tag{34}$$

In the general case when wall slip does occur along the bottom (fixed) plate, the top slip velocity $u_{w2}$ is given by

$$u_{w2}(r) = \left(\beta_1/\beta_2\right)^{1/m_2} u_{w1}^{m_1/m_2}(r) \tag{35}$$

and, therefore, the bottom slip velocity at any radial distance can be found as the solution of

$$\tau_c + \beta_1 u_{w1}^{m_1} = k\left[\dot{\gamma}_a(r) - \frac{\left\{1 + \left(\beta_1/\beta_2\right)^{1/m_2} u_{w1}^{m_1/m_2 - 1}(r)\right\} u_{w1}(r)}{H}\right]^n \tag{36}$$

By means of Eqs. (9) and (27), the above equation can also be written as follows

$$\frac{\beta_1 H^n}{k} u_{w1}^{m_1} = \left[\Omega r - \left\{1 + \left(\beta_1/\beta_2\right)^{1/m_2} u_{w1}^{m_1/m_2 - 1}(r)\right\} u_{w1}(r)\right]^n \text{ - } \Omega_c^n R^n \tag{37}$$

Equation (37) is easily solved using standard methods. Analytical solutions for the slip velocities can be obtained in some special cases. Table 1 shows such solutions for $n = 1$ (Newtonian fluid) and $n = 1/2$. These can also be expressed in terms of the apparent shear rate $\dot{\gamma}_a$ by means of $(\Omega r - \Omega_c R) = H(\dot{\gamma}_a - \dot{\gamma}_{ac})$. Similarly, the rim slip velocities are obtained using $(\Omega - \Omega_c) R = H(\dot{\gamma}_{aR} - \dot{\gamma}_{ac})$. When $\tau_c = 0$, $\Omega_c$ vanishes and wall slip occurs for any finite angular velocity and any radial distance (the first branches of Eqs. (31) and (32) are not relevant).

When the same slip law applies at the two disks, Eq. (34) becomes

$$\tau_c + \beta u_w^m = k\left[\dot{\gamma}_a(r) - \frac{2u_w(r)}{H}\right]^n . \tag{38}$$

The azimuthal velocity, the shear stress, and the apparent flow curve when $\dot{\gamma}_{aR} > \dot{\gamma}_{ac}$ are respectively given by

$$u_\theta(r,z) = \begin{cases} \dot{\gamma}_a(r) z, & 0 \le r \le r_c \\ \left[\dot{\gamma}_a(r) - \dfrac{2u_w(r)}{H}\right] z + u_w(r), & r_c \le r \le R \end{cases} \tag{39}$$

$$\tau_{z\theta}(r) = \begin{cases} k\dot{\gamma}_a^n, & 0 \le r \le r_c \\ k\left[\dot{\gamma}_a(r) - \dfrac{2u_w(r)}{H}\right]^n, & r_c \le r \le R \end{cases} \tag{40}$$

and

$$\tau_R = \begin{cases} k\dot{\gamma}_{aR}^n, & 0 \le \dot{\gamma}_a \le \dot{\gamma}_{ac} \\ k\left(\dot{\gamma}_{aR} - \dfrac{2u_{wR}}{H}\right)^n, & \dot{\gamma}_a > \dot{\gamma}_{ac} \end{cases} \tag{41}$$

Substituting Eq. (40) into Eq. (18) and integrating one finds the second branch of the torque in the case of a power-law fluid:

$$M = 2\pi R^3 \begin{cases} \dfrac{k\dot{\gamma}_{aR}^n}{n+3}, & \dot{\gamma}_a \le \dot{\gamma}_{ac} \\ \dfrac{k}{n+3}\dfrac{\dot{\gamma}_{ac}^{n+3}}{\dot{\gamma}_{aR}^3} + \dfrac{\tau_c}{3}\left(1 - \dfrac{\dot{\gamma}_{ac}^3}{\dot{\gamma}_{aR}^3}\right) + \beta\dfrac{1}{\dot{\gamma}_{aR}^3}\displaystyle\int_{\dot{\gamma}_{ac}}^{\dot{\gamma}_{aR}} u_w^m \dot{\gamma}_a^2 d\dot{\gamma}_a, & \dot{\gamma}_a > \dot{\gamma}_{ac} \end{cases} \tag{42}$$

For the derivative $d\ln M / d\ln\dot{\gamma}_{aR}$ one finds:

$$\frac{d\ln M}{d\ln\dot{\gamma}_{aR}} = \begin{cases} n, & \dot{\gamma}_a \le \dot{\gamma}_{ac} \\ \dfrac{2\pi R^3}{M\dot{\gamma}_{aR}^3}\left\{\dfrac{-3k}{n+3}\dot{\gamma}_{ac}^{n+3} + \dot{\gamma}_{ac}^3\tau_c - 3\beta\displaystyle\int_{\dot{\gamma}_{ac}}^{\dot{\gamma}_{aR}} u_w^m\dot{\gamma}_a^2 d\dot{\gamma}_a + \beta\dot{\gamma}_{aR}^3 u_{wR}^m\right\}, & \dot{\gamma}_a > \dot{\gamma}_{ac} \end{cases} \tag{43}$$

It is straightforward to show that the torque is differentiable (smooth) at $\dot{\gamma}_{aR} = \dot{\gamma}_{ac}$.

Figure 2 illustrates the effect of the gap size $H$ on the apparent flow curve, i.e., the plot of the rim shear stress $\tau_R$ versus the apparent rim shear rate. The initial branch corresponding to no-slip is the same for all gap sizes. The slope of the apparent flow curve changes at $\dot{\gamma}_{ac}$ and is more pronounced with the smaller gap size. In the special case when $n = m = 1$, one finds from Eq. (38) that

$$u_w = \frac{k\dot{\gamma}_a - \tau_c}{\beta + 2k/H} \tag{44}$$

(The above expression is equivalent to the alternative form tabulated in Table 1.) The apparent flow curve in this case is described by

$$\tau_R = \begin{cases} k\dot{\gamma}_{aR}, & 0 \le \dot{\gamma}_a \le \dot{\gamma}_{ac} \\ k\dfrac{\dot{\gamma}_{aR} + 2\tau_c/(\beta H)}{1 + 2k/(\beta H)}, & \dot{\gamma}_a > \dot{\gamma}_{ac} \end{cases} \tag{45}$$

It is clear that the two branches of the apparent flow curve collapse when $\tau_c = 0$ and that the slope of the second branch is lower, as illustrated in Fig. 2.

To illustrate the gap height effect, we caried out numerical experiments on hypothetical power-law fluids in a parallel plate rheometer of radius $R = 25$ mm using three gap heights, i.e., $H = 0.2$, 0.5, and 1 mm and assuming that $k = 1$ Pa s$^n$ and $\beta$=1000 Pa s$^m$ / m$^m$. Figure 3 shows the apparent flow curves obtained in four representative cases. For a Newtonian fluid (Fig. 3a) exhibiting Navier slip, i.e., zero slip yield stress with $m = 1$, only one flow regime is observed. The three apparent flow curves are parallel translations of the no-slip flow curve. As expected, slip effects become more pronounced as the gap height is reduced. When the slip yield stress is nonzero, two flow regimes are observed, as in Figs. 3b-c. Below the critical apparent shear rate $\dot{\gamma}_{ac}$, all flow curves coincide with the no-slip flow curve and then exhibit a plateau which is more visible and longer for smaller gap heights. At higher shear rates the flow curves approach asymptotically their counterparts for non-zero slip yield stress. It should be noted that for a shear thinning fluid (Fig. 3c) the flow curves are not parallel, but they approach asymptotically the no-slip flow curve. Finally, the plateau region after $\dot{\gamma}_{ac}$ is enhanced when the slip exponent $m$ is greater than unity (Fig. 3d).

To visualize the effect of the apparent shear rate (i.e., the angular velocity $\Omega$) on the velocity distribution, let us first consider the variation of the dimensionless rim slip velocity $u_{wR}/(\Omega R)$ for all the cases of Fig. 3. In the case of a Newtonian fluid with zero slip yield stress, $u_{wR}/(\Omega R)$ is constant and is reduced as the gap size is increased (Fig. 4a). With a finite slip yield stress, the relative slip velocity is zero below the critical apparent shear rate $\dot{\gamma}_{aRc}$ and then increases rapidly to converge asymptotically to its zero-slip-yield-stress counterparts (Fig. 4b). However, this final plateau is observed only with Newtonian liquids when the slip exponent $m$ is unity. For lower values of $n$ (Fig. 4c) or higher values of $m$ (Fig. 4d) the dimensionless rim velocity reaches a maximum and then is reduced rapidly as the apparent shear rate is increases. In all cases, the relative slip velocity is reduced with the gap size.

The effect of the apparent shear rate on the contours of the dimensionless velocity $u_\theta/(\Omega R)$ when the slip yield stress is non-zero is illustrated in Figs. 5 and 6, for a Newtonian ($n = 1$) and a power-law fluid ($n = 0.5$). The corresponding no-slip solutions hold up to the critical angular velocity $\Omega_c$, and as $\Omega$ is increased slip is observed only for $r_c \le r \le R$, where the radius $r_c$ vanishes asymptotically. It can also be observed that, due to wall slip, the dimensionless velocity is reduced on the upper disk and increases on the lower disk as the apparent shear rate is increased approaching asymptotically the solid-body motion. The apparent shear rates corresponding to the contour plots of Figs. 5 and 6 are shown in Fig. 7 along with the corresponding apparent flow curves.

The gap-size effects on the plots of the torque versus the apparent rim shear rate are essentially the same as those on the flow curves. This is easily deduced by comparing the torque plots in Fig. 8 with the corresponding flow curves shown in Fig. 3. When $\tau_c$ is zero, only one flow regime is observed (given by the second branch of Eq. (42)), and $M$ is everywhere gap dependent (Fig. 8a). When $\tau_c$ is non-zero, two regimes are observed (defined by $\dot{\gamma}_{ac}$) and only the first branch of the flow curve is gap-independent; see Figs. 8b-d. Interestingly, the rheological parameters and the slip yield stress can be determined from the first branch of the plot of $\ln M$ vs $\ln \dot{\gamma}_{aR}$: the power-law exponent $n$ is simply the slope of this branch and the consistency index $k$ can be determined from $M_c = M(\dot{\gamma}_{ac})$:

$$k = \frac{(n+3)M_c}{2\pi R^3 \dot{\gamma}_{ac}^n} \tag{46}$$

Finally, the slip yield stress can be determined by means of $\tau_c = k\dot{\gamma}_{ac}^n$. The other slip parameters, i.e., $\beta$ and $m$ can be determined from the second branch, which is gap dependent. If the fluid is not shear-thinning, $M$ tends asymptotically to its Navier-slip counterpart (Fig. 8b). Otherwise, it approaches asymptotically the gap-size-independent no-slip curve (Fig. 8c). As illustrated in Fig. 8d, when the slip exponent $m$ is increased above unity, the second branch of the torque becomes flatter.

## 4. Torsional flow of a Herschel-Bulkley fluid

In this section we consider the torsional flow of a Herschel-Bulkley fluid, assuming that the same slip law with non-zero slip yield stress applies along the two plates and that $0 < \tau_c < \tau_y$. Therefore, three distinct regimes are encountered as the rim shear stress $\tau_R$ is increased.

(i) If $\tau_R \le \tau_c$ the exerted torque is not sufficient to rotate the disk and the material remains unyielded.

(ii) When $\tau_c < \tau_R \le \tau_y$ the material is still unyielded, but exhibits slip and rotates as a solid at half the angular velocity of the disk

$$u_\theta(r) = \frac{\Omega r}{2} \tag{47}$$

and the shear stress is given by

$$\tau_{z\theta}(r) = \tau_c + \beta\left(\frac{\Omega r}{2}\right)^m = \tau_c + \beta\left(\frac{H\dot{\gamma}_a}{2}\right)^m, \quad 0 \le r \le R \tag{48}$$

Since $u_w(R) = \Omega R/2 = \left(H\dot{\gamma}_{aR}/2\right)$, the rim shear stress,

$$\tau_R = \tau_c + \beta\left(\frac{H\dot{\gamma}_{aR}}{2}\right)^m, \tag{49}$$

Is gap-size dependent. The critical angular speed $\Omega_y$ at which the material starts yielding is obtained by demanding $\tau_R = \tau_y$, which gives

$$\Omega_y = \frac{2}{R}\left(\frac{\tau_y - \tau_c}{\beta}\right)^{1/m} \tag{50}$$

Hence, the critical shear rate at the rim is

$$\dot{\gamma}_{aRy} = \frac{2}{H}\left(\frac{\tau_y - \tau_c}{\beta}\right)^{1/m} \tag{51}$$

and the critical rim slip velocity is

$$u_{wRy} = \left(\frac{\tau_y - \tau_c}{\beta}\right)^{1/m} \tag{52}$$

We observe that $\dot{\gamma}_{aRy}$ decreases with the gap size $H$ in agreement with experimental observations.

(iii) When $\tau_R > \tau_y$, the material exhibits slip everywhere but yields only in the annulus $r_y \le r \le R$, where $r_y$ is the critical radius at which $\tau_{z\theta} = \tau_y$. Hence, for $0 \le r \le r_y$ the material rotates unyielded following Eq. (47) and the shear stress is given by

$$\tau_{z\theta}(r) = \tau_c + \beta\left(\frac{\Omega r}{2}\right)^m, \quad 0 \le r \le r_y \tag{53}$$

By demanding that $\tau_{z\theta} = \tau_y$, one finds that

$$r_y = \frac{2}{\Omega}\left(\frac{\tau_y - \tau_c}{\beta}\right)^{1/m} \tag{54}$$

The slip velocity for $r_y \le r \le R$ is found by solving

$$\tau_{z\theta} = \tau_c + \beta u_w^m = \tau_y + k\left[\dot{\gamma}_a(r) - \frac{2u_w(r)}{H}\right]^n \tag{55}$$

which generalizes Eq. (38). The azimuthal velocity is given by

$$u_\theta(r,z) = \begin{cases} \dfrac{\Omega r}{2} = \dfrac{H\dot{\gamma}_a(r)}{2}, & 0 \le r \le r_y \\ \left[\dot{\gamma}_a(r) - \dfrac{2u_w(r)}{H}\right] z + u_w(r), & r_y \le r \le R \end{cases} \tag{56}$$

Then, the apparent flow curve is described by

$$\tau_R = \begin{cases} \tau_c + \beta\left(\dfrac{H\dot{\gamma}_{aR}}{2}\right)^m, & 0 \le \dot{\gamma}_a \le \dot{\gamma}_{aRy} \\ \tau_y + k\left(\dot{\gamma}_{aR} - \dfrac{2u_{wR}}{H}\right)^n, & \dot{\gamma}_a > \dot{\gamma}_{aRy} \end{cases} \tag{57}$$

The effect of the gap size on the apparent flow curve is illustrated in Fig. 9. As predicted by Eq. (51), the critical apparent shear rate $\dot{\gamma}_{aRy}$ is also gap-dependent. This fact can be exploited in order to determine the slip parameters $\beta$ and $m$ from experimental data obtained using different gap sizes.

The torque in the two flow regimes is found to be:

$$M = 2\pi R^3 \begin{cases} \dfrac{\tau_c}{3} + \dfrac{\beta H^m}{2^m (m+3)} \dot{\gamma}_{aR}^m, & \dot{\gamma}_a \le \dot{\gamma}_{aRy} \\ \dfrac{\tau_c}{3}\left[1 - \dfrac{3}{m+3}\left(\dfrac{\dot{\gamma}_{aRy}}{\dot{\gamma}_{aR}}\right)^3\right] + \dfrac{\tau_y}{m+3}\left(\dfrac{\dot{\gamma}_{aRy}}{\dot{\gamma}_{aR}}\right)^3 + \dfrac{\beta}{\dot{\gamma}_{aR}^3}\int_{\dot{\gamma}_{aRy}}^{\dot{\gamma}_{aR}} u_w^m \dot{\gamma}_a^2 d\dot{\gamma}_a, & \dot{\gamma}_a > \dot{\gamma}_{aRy} \end{cases} \tag{58}$$

Quan et al. (2023) noted that an approximate analytical expression for the integral term in the second branch of Eq. (58) can be obtained by assuming that in the yield region $r_y \le r \le R$ the wall slip contribution to shear stress is negligible, that is

$$\tau_{z\theta} = \tau_c + \beta u_w^m = \tau_y + k\left[\dot{\gamma}_a(r) - \frac{2u_w(r)}{H}\right]^n \simeq \tau_y + k\dot{\gamma}_a^n(r), \quad r_y \le r \le R \tag{59}$$

Under this assumption one gets:

$$M = 2\pi R^3 \left\{ \frac{\tau_c}{3}\frac{\dot{\gamma}_{aRy}^3}{\dot{\gamma}_{aR}^3} + \frac{\beta H^m}{2^m (m+3)}\frac{\dot{\gamma}_{aRy}^{m+3}}{\dot{\gamma}_{aR}^3} + \frac{\tau_y}{3}\left(1 - \frac{\dot{\gamma}_{aRy}^3}{\dot{\gamma}_{aR}^3}\right) + \frac{k}{n+3}\dot{\gamma}_{aR}^n\left(1 - \frac{\dot{\gamma}_{aRy}^{n+3}}{\dot{\gamma}_{aR}^{n+3}}\right)\right\}, \quad \dot{\gamma}_a > \dot{\gamma}_{aRy} \tag{60}$$

A limitation of the above approximation, however, is that the curve of $\ln M$ vs. $\ln \dot{\gamma}_{aR}$ is not differentiable at $\dot{\gamma}_a = \dot{\gamma}_{aR}$.

Some special analytical solutions for the slip velocity and the rim shear stress are tabulated in Table 2. Equivalent expressions for the flow curves have also been obtained in Georgiou (2021) for the case of simple shear flow. It is straightforward to express these results in terms of the apparent shear rate, by means of $\left(\Omega r - \Omega_y R\right) = H\left(\dot{\gamma}_a - \dot{\gamma}_{aRy}\right)$. For example, when $n = m = 1$, the solution of Eq. (55) is

$$u_w = \frac{k\dot{\gamma}_a + \tau_y - \tau_c}{\beta + 2k/H} \tag{61}$$

by means of which one gets the following expression for the apparent flow curve:

$$\tau_R = \begin{cases} \tau_c + \beta\left(\dfrac{H\dot{\gamma}_{aR}}{2}\right), & 0 \le \dot{\gamma}_a \le \dot{\gamma}_{aRc} \\ \dfrac{k\dot{\gamma}_{aR} + \tau_y + 2k\tau_c/(\beta H)}{1 + 2k/(\beta H)}, & \dot{\gamma}_a > \dot{\gamma}_{aRc} \end{cases} \tag{62}$$

Referring to Fig. 9, the slopes of the two branches are $\beta H/2$ and $1/[1/k + 2/(\beta H)]$, respectively. Obviously, both slopes increase with the gap size. Extrapolating the second branch to $\dot{\gamma}_{aR} = 0$ leads to an expression relating the slip parameters with the rheological ones. Setting $\tau_c = 0$ leads to the zero-slip-yield-stress dashed curves of Fig. 9. In case $\tau_c = \tau_y$, then $\dot{\gamma}_{aRy} = 0$ and the apparent flow curve consists of a single branch the slope of which is lower than that corresponding to the no-slip apparent flow curve.

Figure 9 shows representative apparent flow curves obtained from numerical experiments on Herschel-Bulkley fluids exhibiting wall slip. To illustrate the effect of the gap height we consider again a real parallel plate rheometer with $H = 0.2$, 0.5, and 1 mm and assume that $\tau_y = 2$ Pa and $\tau_c$ =0 or 0.5 Pa (the slip yield stress is lower than the yield stress). Two flow regimes are observed. The first one corresponds to solid-body rotation (the material is unyielded) and in the second the material is partially yielded far from the symmetry axis. If the slip yield stress is zero (Fig. 10a), the shear stress increases rapidly with the apparent shear rate and at the critical apparent shear rate where the yield stress is reached the flow curve passes through a plateau and eventually it increases again rapidly approaching asymptotically the no-slip flow curve. It is clear that the stress corresponding to the plateau of the flow curve provides an estimate for the yield stress of the material. Moreover, the values of the critical apparent shear rate $\dot{\gamma}_{aRy}$ at which the plateau starts for the different gap heights can be used to obtain initial estimates of the slip parameters $\beta$ and $m$, by means of Eq. (51). In the case of non-zero slip yield stress (Figs. 10b-d), the apparent flow curve is characterized by two plateaux, the levels of which provide good estimates of $\tau_c$ and $\tau_y$. The first plateau, which is below the material yield stress, is referred to as 'dynamic yield stress' (Ewoldt et al., 2015).

When slip yield stress is nonzero, gap-size effects appear to be important only in an intermediate range of apparent shear rates, which includes the critical apparent shear rate $\dot{\gamma}_{aRy} = 0$. It can be observed in Figs. 10b-d that the apparent flow curves for different gap sizes tend to merge at low and high apparent shear rates, in agreement with experimental data on colloidal suspensions (Moud et al., 2021; Ewoldt et al., 2015).

The variation of the dimensionless rim slip velocity with the apparent shear rate in the case of yield stress fluids is quite different from that of the power-law fluids discussed in Fig. 4. This is due to the presence of the solid-body rotation regime where the slip velocity is at maximum, i.e., $u_{wR}/(\Omega R) = 1/2$. As illustrated in Fig. 11, where the slip velocities for the cases considered in Fig. 10 are plotted, beyond the critical shear rate $\dot{\gamma}_{aRc}$, the slip velocity reduces rapidly and then tends asymptotically to a constant value, which increases as the gap size is reduced. The reduction of the dimensionless slip velocity is much faster with shear thinning fluids (Fig. 11c) and no plateau is reached when $n<1$. This reduction is also slower when the slip exponent $m$ is greater than unity (Fig. 11d).

The yield radius $r_y$ in the yielding regime ($\dot{\gamma}_a \geq \dot{\gamma}_{aRy}$) is inversely proportional to the gap height, as dictated by Eq. (54), which can also be written in the following form:

$$\frac{r_y}{R} = \frac{2}{\dot{\gamma}_{aR} H} \left( \frac{\tau_y - \tau_c}{\beta} \right)^{1/m} \tag{63}$$

A representative plot showing the effect of the gap height $H$ on the yield radius is provided in Fig. 12a, where the parameter values are those of the flow curves depicted in Fig. 10b. It is clear that the rim slip velocity is reduced as more material becomes yielded. The variation of the rim slip velocity with the yield radius is shown in Fig. 12b. The maximum of the rim slip velocity (1/2) obviously occurs when the fluid is unyielded ($r_y/R$ is unity).

Figures 13 and 14 show the dimensionless velocity contours at different apparent shear rates for a Bingham plastic and a Herschel-Bulkley fluid with $n = 0.5$, respectively, in the case of non-zero slip yield stress. Below the critical apparent shear rate $\dot{\gamma}_{aRy}$, the material rotates as a solid. Above $\dot{\gamma}_{aRy}$, the material yields only in the region $r_y \leq r \leq R$, which causes the relative velocity $u_\theta/(\Omega R)$ to increase at the top plate and to reduce at the lower plate, since the relative slip velocity is reduced. This effect is opposite to that observed with the power-law fluids in Figs. 5 and 6. As a result, the velocity contours in the yielded region are bended to the right and the relative velocity at the rim increases. The values of the apparent shear rates used for the contours plots of Figs. 13 and 14, are shown in Fig. 15 along with the corresponding apparent flow curves.

Finally, Figure 16 shows the calculated torques versus the apparent rim shear rates obtained for the three gap sizes and the geometric and material parameters used for the flow curves shown in Fig. 10. As already noted both branches of the resulting curves are gap dependent in contrast to the power-law fluids. When the slip yield stress is non zero the curves for different gap sizes are flat initially and essentially coincide. Another difference from the power-law-fluid case is that the critical shear rate $\dot{\gamma}_{aRy}$ is gap dependent (as dictated by Eq. (51). In fact, estimates of the yield stresses $\tau_c$ and $\tau_y$ as well as the slip coefficient $\beta$ and the slip exponent $m$ can be determined from torque data obtained only in the first (unyielded) regime. More specifically, the slip yield stress is directly calculated from the initial torque plateau corresponding say to $M_0$, $\tau_c = 3M_0 / (2\pi R^3)$. Then, the yield stress and the slip exponent can be found from the critical torque $M_y = M(\dot{\gamma}_{aRy})$ and the slope of the left branch of the $\ln M$ vs $\ln \dot{\gamma}_{aR}$ curve at $\dot{\gamma}_{aRy}$, i.e., by solving the system

$$M_y = \frac{2\pi R^3 (m\tau_c + 3\tau_y)}{3(m+3)} \quad \text{and} \quad \frac{d\ln M}{d\ln \dot{\gamma}_{aR}}(\dot{\gamma}_{aRy}^-) = \frac{3m(\tau_y - \tau_c)}{3\tau_y + m\tau_c} \tag{64}$$

It should be pointed out that the slope of the torque curve at $\dot{\gamma}_{aRy}$ is independent of the gap size. Finally, the slip coefficient can be determined by means of Eq. (51). Data in the yielding regime are required in order to determine the other rheological parameters ($k$ and $n$).

**5. Conclusions**

The torsional parallel plate flow of Herschel-Bulkley fluids has been studied assuming that wall slip with nonzero slip yield stress occurs at both plates. The slip yield stress was taken to be lower than the yield stress and the resulting flow regimes have been identified. The velocity and stress fields are obtained by means of explicit analytical expressions in terms of the slip velocity, which is calculated numerically in the general case. The gap-size effects on the apparent flow curve and the torque have been demonstrated for both power-law and Herschel-Bulkley flows. Analytical solutions for certain combinations of the power-law and slip exponents are provided and the effects of wall slip on the two-dimensional flow field have been discussed.

## Data availability

The data that support the findings of this study are available from the corresponding author upon reasonable request.

**Table 1**. Slip velocities for Newtonian $(n=1)$ and power-law $(n=1/2)$ fluids

| | |
|---|---|
| **Newtonian fluid** <br> $\tau = k\dot{\gamma}$ <br> $(\tau_y = 0, \quad n = 1)$ | Slip only in $r_c = \frac{H\tau_c}{k\Omega} \le r \le R$ when $\Omega > \Omega_c = \frac{H\tau_c}{kR}$ |
| $m_1 = m_2 = 1$ | $u_{w1} = \frac{\Omega r - \Omega_c R}{1 + \beta_1 H / k + \beta_1 / \beta_2}$, $\quad u_{w2} = \frac{\beta_1}{\beta_2} u_{w1}$ <br><br> When $\beta_1 = \beta_2 = \beta,\ u_w = \frac{\Omega r - \Omega_c R}{2 + \beta H / k}$ |
| $m_1 = m_2 = 2$ | $u_{w1} = \frac{2\left(\Omega r - \Omega_c R\right)}{1 + \sqrt{\beta_1 / \beta_2} + \sqrt{\left(1 + \sqrt{\beta_1 / \beta_2}\right)^2 + \frac{4H\beta_1}{k}\left(\Omega r - \Omega_c R\right)}}$, $\quad u_{w2} = \sqrt{\beta_1 / \beta_2}\, u_{w1}$ <br><br> When $\beta_1 = \beta_2 = \beta,\ u_w = \frac{\Omega r - \Omega_c R}{1 + \sqrt{1 + \frac{H\beta}{k}\left(\Omega r - \Omega_c R\right)}}$ |
| $m_1 = m_2 = 1/2$ | $u_{w1} = \frac{4\left(\frac{k}{\beta_1 H}\right)^2 \left(\Omega r - \Omega_c R\right)^2}{\left[1 + \sqrt{1 + 4\left[1 + (\beta_1 / \beta_2)^2\right]\left(\frac{k}{\beta_1 H}\right)^2 \left(\Omega r - \Omega_c R\right)}\right]^2}$, $\quad u_{w2} = \left(\frac{\beta_1}{\beta_2}\right)^2 u_{w1}$ <br><br> When $\beta_1 = \beta_2 = \beta,\ u_w = \frac{4\left(\frac{k}{\beta H}\right)^2 \left(\Omega r - \Omega_c R\right)^2}{\left[1 + \sqrt{1 + 8\left(\frac{k}{\beta H}\right)^2 \left(\Omega r - \Omega_c R\right)}\right]^2}$ |
| **Power-law fluid** <br> $\tau = k\dot{\gamma}^{1/2}$ <br> $(\tau_y = 0, \quad n = 1/2)$ | Slip only in $r_c = \frac{H}{\Omega}\left(\frac{\tau_c}{k}\right)^2 \le r \le R$ when $\Omega > \Omega_c = \frac{H}{R}\left(\frac{\tau_c}{k}\right)^2$ |
| $m_1 = m_2 = 1$ | $u_{w1} = \frac{2\left(\Omega r - \Omega_c R\right)}{\left(2\sqrt{\Omega_c RH}\, \beta_1/k + 1 + \beta_1/\beta_2\right)\left[1 + \sqrt{1 + \frac{4\beta_1^2 H\left(\Omega r - \Omega_c R\right)}{k^2\left(2\sqrt{\Omega_c RH}\, \beta_1/k + 1 + \beta_1/\beta_2\right)^2}}\right]}$, $\quad u_{w2} = \left(\frac{\beta_1}{\beta_2}\right) u_{w1}$ <br> The solution for $\beta_1 = \beta_2 = \beta$ is easily deduced. |
| $m_1 = m_2 = 1/2$ | $u_{w1} = \frac{\frac{k^2 \Omega_c R}{\beta_1^2 H}\left[\sqrt{1 + \left[1 + \frac{k^2}{H}\left(\frac{1}{\beta_1^2} + \frac{1}{\beta_2^2}\right)\right]\left(\frac{\Omega r}{\Omega_c R} - 1\right)} - 1\right]^2}{\left[1 + \frac{k^2}{H}\left(\frac{1}{\beta_1^2} + \frac{1}{\beta_2^2}\right)\right]^2}$, $\quad u_{w2} = \left(\frac{\beta_1}{\beta_2}\right)^2 u_{w1}$ <br> The solution for $\beta_1 = \beta_2 = \beta$ is easily deduced. |

**Table 2**. Slip velocities and flow curves for Bingham $(n=1)$ and Herschel-Bulkley $(n=1/2)$ fluids

| | |
|---|---|
| $n=m=1$ <br> $\tau=\tau_y+k\dot{\gamma}$ <br> $\tau_w=\tau_c+\beta u_w$ | $\Omega_y=\frac{2}{R}\left(\frac{\tau_y-\tau_c}{\beta}\right),\ \dot{\gamma}_{aRy}=\frac{2}{H}\left(\frac{\tau_y-\tau_c}{\beta}\right),\ r_0=\frac{2}{\Omega}\left(\frac{\tau_y-\tau_c}{\beta}\right),\ \Omega>\Omega_y$ <br> The slip velocity is given by <br> $u_w(r)=\begin{cases}\frac{\Omega r}{2}, & 0\le r\le r_0\\ \frac{k\Omega r/H+\tau_y-\tau_c}{\beta+2k/H}, & r_0\le r\le R\end{cases}$ <br> $u_\theta(r,z)=\begin{cases}\frac{\Omega r}{2}, & 0\le r\le r_0\\ \frac{k\Omega r/H+\tau_y-\tau_c}{\beta+2k/H}\left(1-\frac{2z}{H}\right)+\frac{\Omega rz}{H}, & r_0<r\le R\end{cases},\quad \Omega>\Omega_y$ <br> $\tau_R=\begin{cases}\tau_c+\beta\left(\frac{H\dot{\gamma}_{aR}}{2}\right), & \dot{\gamma}_{aR}\le\dot{\gamma}_{aRy}\\ \frac{k\dot{\gamma}_{aR}+\tau_y+2k\tau_c/(\beta H)}{1+2k/(\beta H)}, & \dot{\gamma}_{aR}>\dot{\gamma}_{aRy}\end{cases}$ |
| $n=1,\ \ m=2$ <br> $\tau=\tau_y+k\dot{\gamma}$ <br> $\tau_w=\tau_c+\beta u_w^2$ | $\Omega_y=\frac{2}{R}\left(\frac{\tau_y-\tau_c}{\beta}\right)^{1/2},\ \dot{\gamma}_{aRy}=\frac{2}{H}\left(\frac{\tau_y-\tau_c}{\beta}\right)^{1/2},\ r_0=\frac{2}{\Omega}\left(\frac{\tau_y-\tau_c}{\beta}\right)^{2},\ \Omega>\Omega_y$ <br> $\tau_R=\begin{cases}\tau_c+\beta\left(\frac{H\dot{\gamma}_{aR}}{2}\right)^2, & \dot{\gamma}_{aR}\le\dot{\gamma}_{aRy}\\ \tau_y+k\left(\dot{\gamma}_{aR}-\frac{2u_{wR}}{H}\right), & \dot{\gamma}_{aR}>\dot{\gamma}_{aRy}\end{cases}$ <br> where <br> $u_{wR}=\frac{1}{\beta H}\left[\sqrt{k^2+\beta H^2(\tau_y-\tau_c)+\beta H^2k\dot{\gamma}_{aR}}-k\right]$ |
| $n=1/2,\ \ m=1$ <br> $\tau=\tau_y+k\dot{\gamma}^{1/2}$ <br> $\tau_w=\tau_c+\beta u_w$ | $\Omega_y=\frac{2}{R}\left(\frac{\tau_y-\tau_c}{\beta}\right),\ \dot{\gamma}_{aRy}=\frac{2}{H}\left(\frac{\tau_y-\tau_c}{\beta}\right),\ r_0=\frac{2}{\Omega}\left(\frac{\tau_y-\tau_c}{\beta}\right),\ \Omega>\Omega_y$ <br> $\tau_R=\begin{cases}\tau_c+\frac{\beta H}{2}\dot{\gamma}_{aR}, & \dot{\gamma}_{aR}\le\dot{\gamma}_{aRy}\\ \tau_y+k\left(\dot{\gamma}_{aR}-\frac{2u_{wR}}{H}\right)^{1/2}, & \dot{\gamma}_{aR}>\dot{\gamma}_{aRy}\end{cases}$ <br> where <br> $u_{wR}=\frac{-k^2+\beta H(\tau_y-\tau_c)+k\sqrt{k^2-2\beta H(\tau_y-\tau_c)+\beta^2H^2\dot{\gamma}_{aR}}}{\beta^2H}$ |

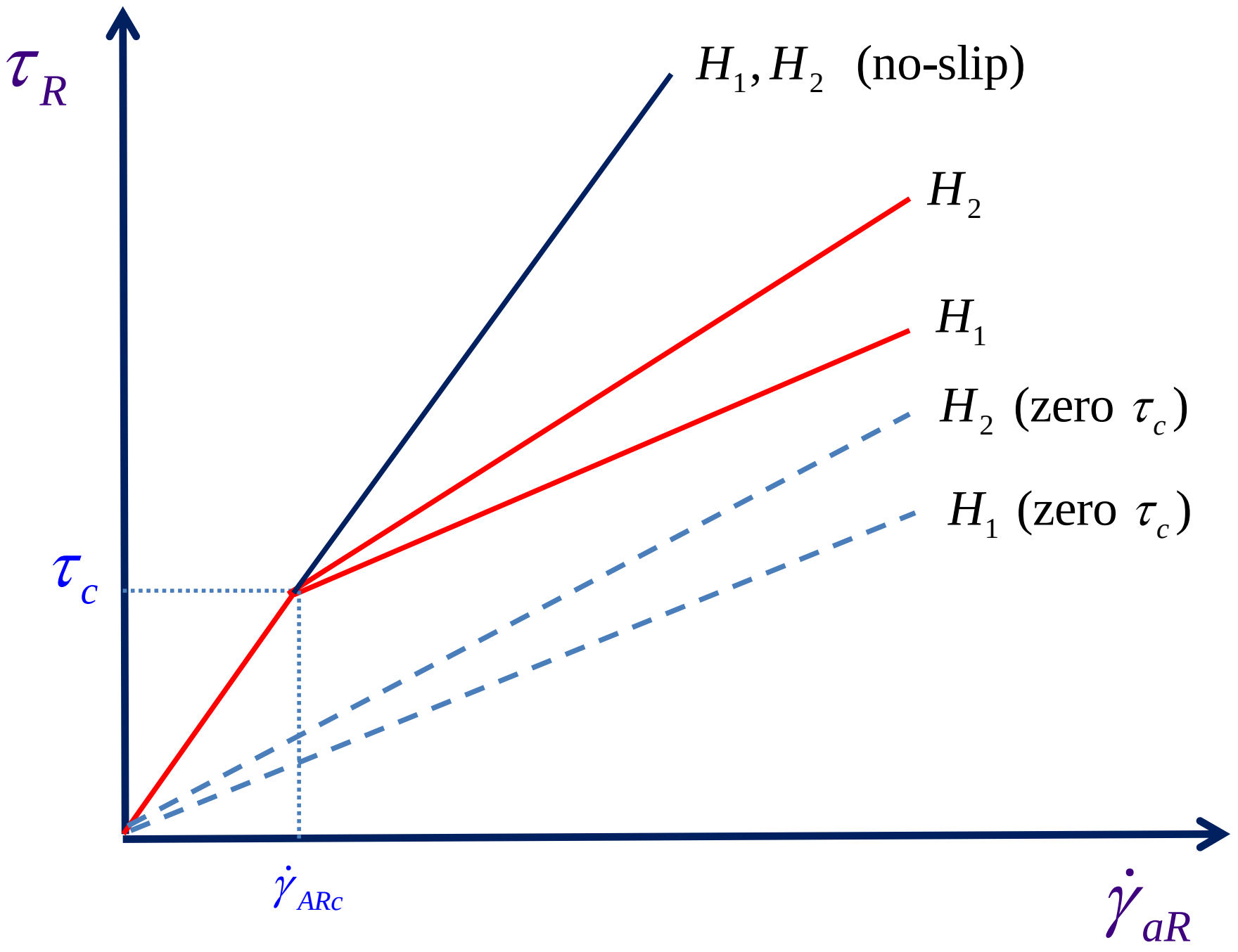

**Figure 2**. Sketch of the gap-size effect ($H_2 > H_1$) on the apparent flow curve of a non-viscoplastic (e.g., power-law) fluid in the presence of wall slip with non-zero slip yield stress. The dashed lines show the apparent flow curves in the case of zero slip yield stress.

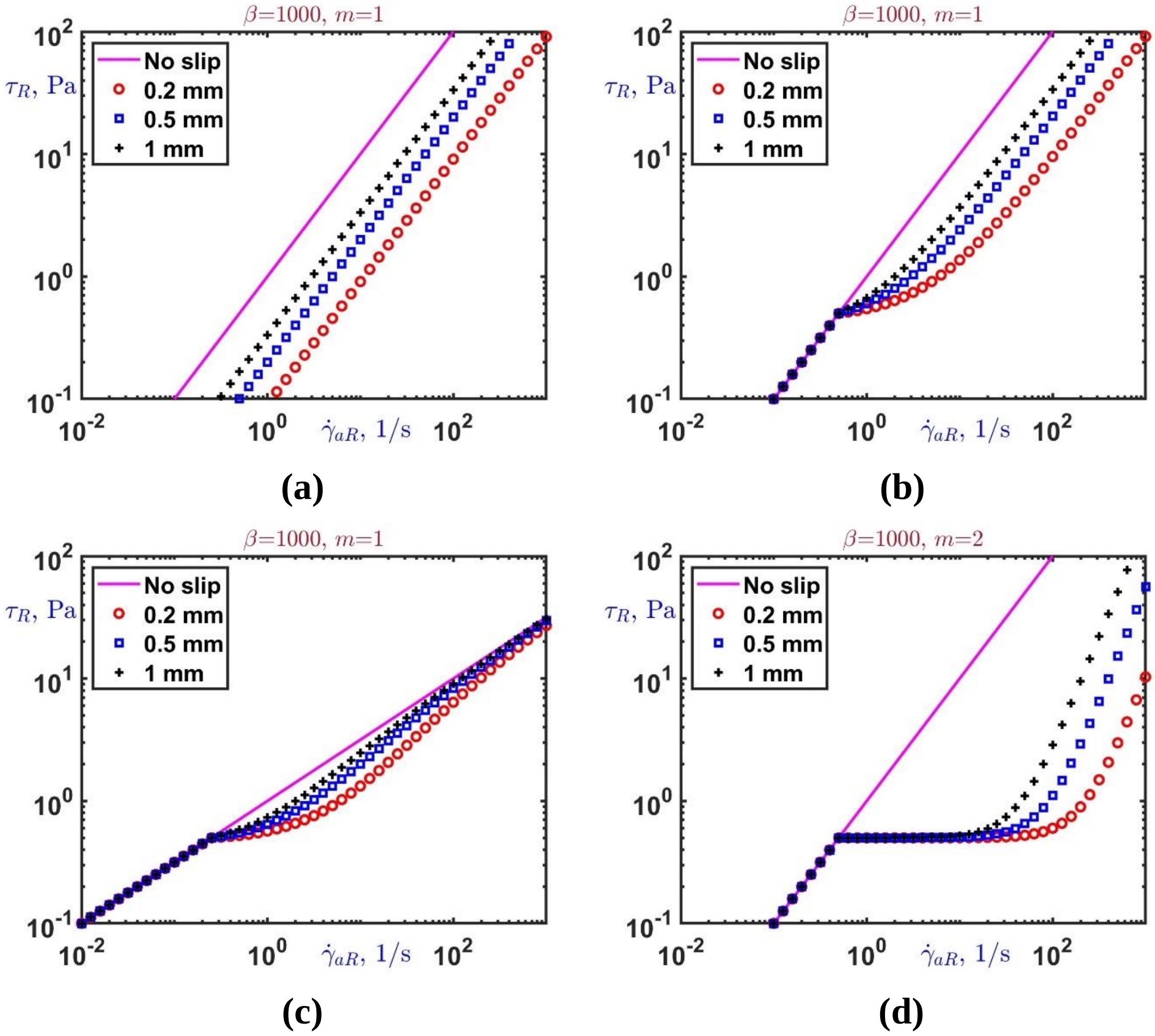

**Figure 3**. Representative apparent flow curves of power-law fluids obtained with three gap heights, $H = 0.2$ mm (o), 0.5 mm (□), and 1 mm (∗), $R = 25$ mm, $k = 1$ Pa s$^n$ and $\beta$=1000 Pa s$^m$ / m$^m$: (a) $n = 1,\ \tau_c = 0,\ \ m = 1$ (Newtonian fluid, zero slip yield stress); (b) $n = 1,\ \tau_c = 0.5$ Pa, $m = 1$ (Newtonian fluid, non-zero slip yield stress); (c) $n = 0.5,\ \tau_c = 0.5$ Pa, $m = 1$ (power-law fluid, non-zero slip yield stress); (d) $n = 1,\ \tau_c = 0.5$ Pa, $m = 2$ (Newtonian fluid, non-zero slip yield stress). The solid line is the flow curve in the case of no-slip.

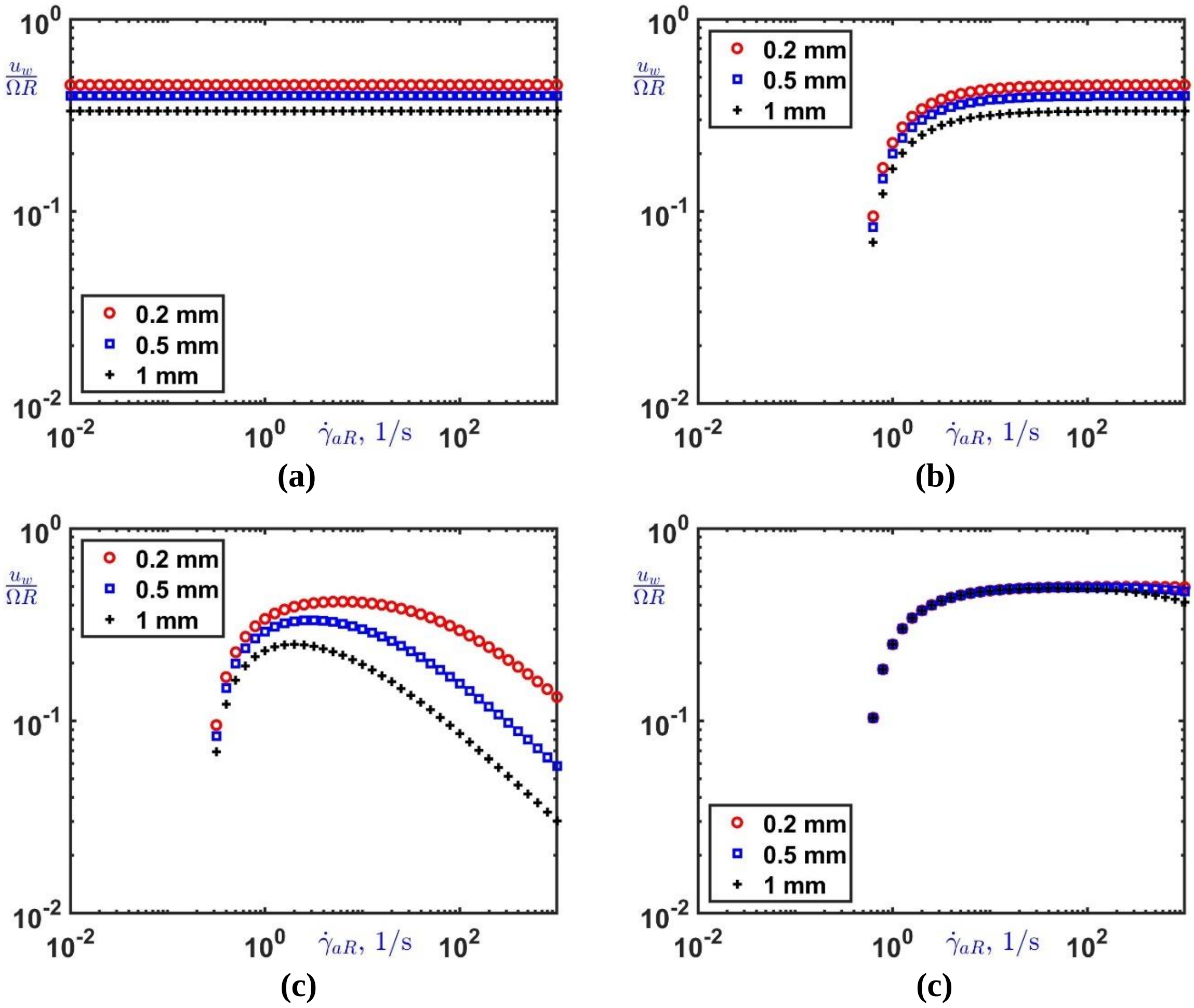

**Figure 4**. Dimensionless rim slip velocities of power-law fluids obtained with three gap heights ($H = 0.2,\ 0.5$ and $1$ mm), $R = 25$ mm, $k = 1$ Pa s$^n$ and $\beta$=1000 Pa s$^m$ / m$^m$: (a) $n = 1,\ \tau_c = 0,\ \ m = 1$ (Newtonian fluid, zero slip yield stress); (b) $n = 1,\ \tau_c = 0.5$ Pa, $m = 1$ (Newtonian fluid, non-zero slip yield stress); (c) $n = 0.5,\ \tau_c = 0.5$ Pa, $m = 1$ (power-law fluid, non-zero slip yield stress); (d) $n = 1,\ \tau_c = 0.5$ Pa, $m = 2$ (Newtonian fluid, non-zero slip yield stress).

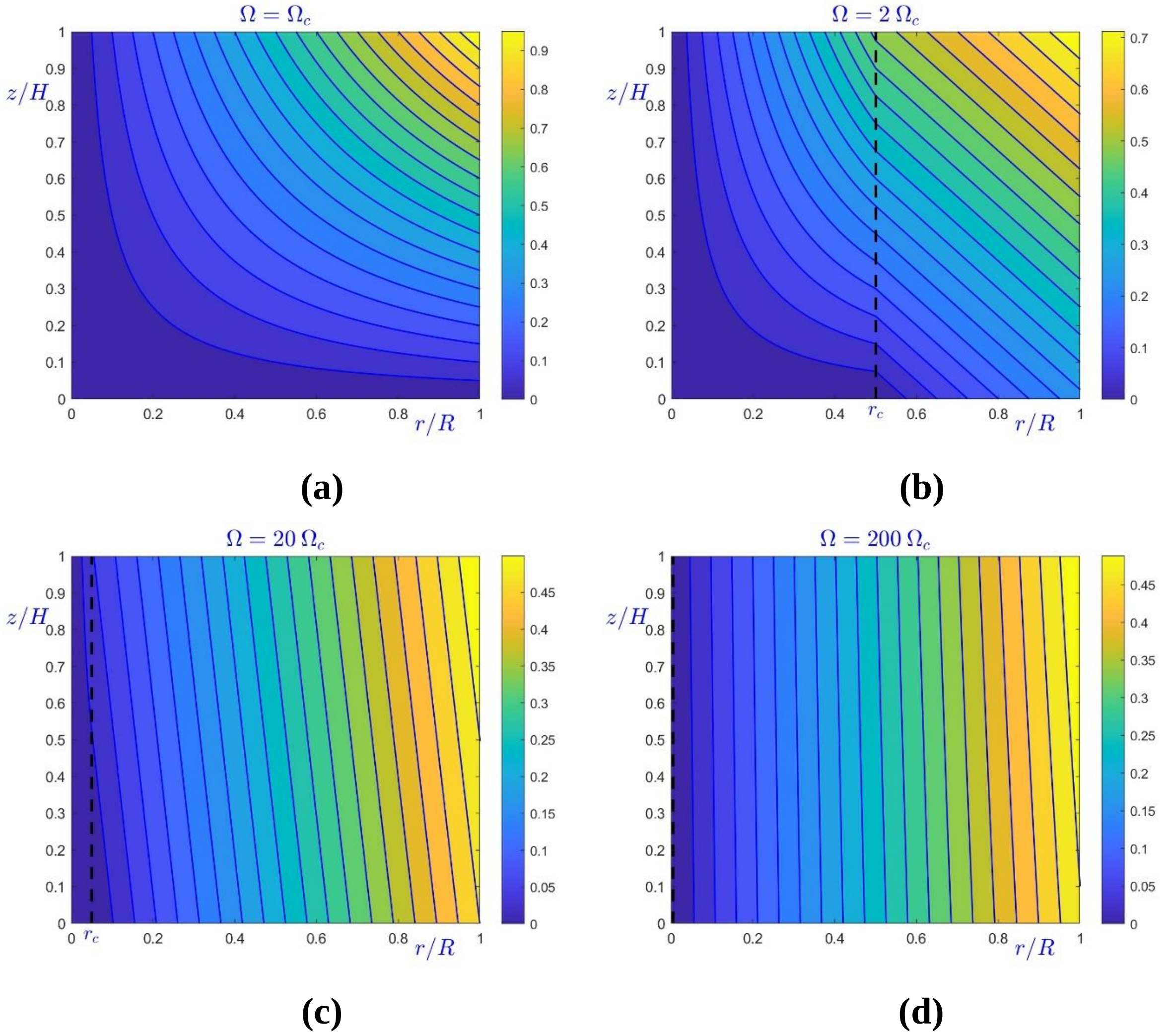

**Figure 5**. Contours of the dimensionless velocity $u_\theta / (\Omega R)$ for various angular velocities in the case of Newtonian flow with non-zero slip yield stress, i.e., $R = 25$ mm, $H = 1$ mm, $k = 1$ Pa s, $\tau_c = 0.5$ Pa, $\beta$=1000 Pa s / m, and $m = 1$: (a) $\Omega = \Omega_c$; (b) $\Omega = 2\Omega_c$; (c) $\Omega = 20\Omega_c$; (d) $\Omega = 200\Omega_c$. The critical angular velocity and apparent shear rate are $\Omega_c = 0.02\ \mathrm{s}^{-1}$ and $\dot{\gamma}_{aRc} = 0.5\ \mathrm{s}^{-1}$. The dashed line denotes the radius $r_c$, below which no slip occurs.

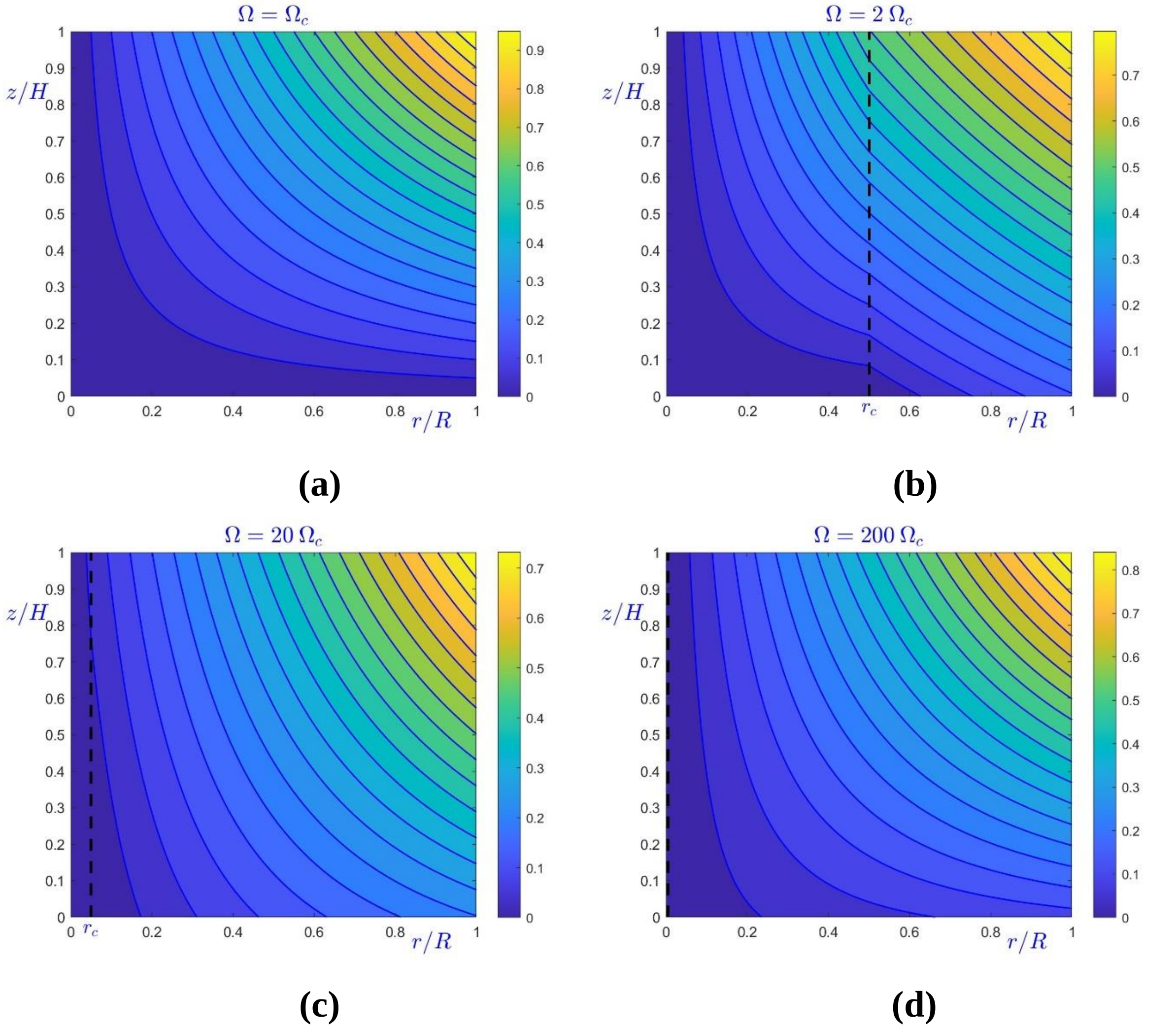

**Figure 6**. Contours of the dimensionless velocity $u_\theta / (\Omega R)$ for various angular velocities in the case of the flow of a power-law fluid with non-zero slip yield stress, i.e., $R = 25\ \mathrm{mm}$, $H = 1\ \mathrm{mm}$, $k = 1\ \mathrm{Pa\ s}$, $n = 1$, $\tau_c = 0.5\ \mathrm{Pa}$, $\beta = 1000\ \mathrm{Pa\ s}^m / \mathrm{m}^m$, and $m = 2$: (a) $\Omega = \Omega_c$; (b) $\Omega = 2\Omega_c$; (c) $\Omega = 20\Omega_c$; (d) $\Omega = 200\Omega_c$. The critical angular velocity and apparent shear rate are $\Omega_c = 0.01\ \mathrm{s}^{-1}$ and $\dot{\gamma}_{aRc} = 0.25\ \mathrm{s}^{-1}$. The dashed line denotes the radius $r_c$, below which no slip occurs.

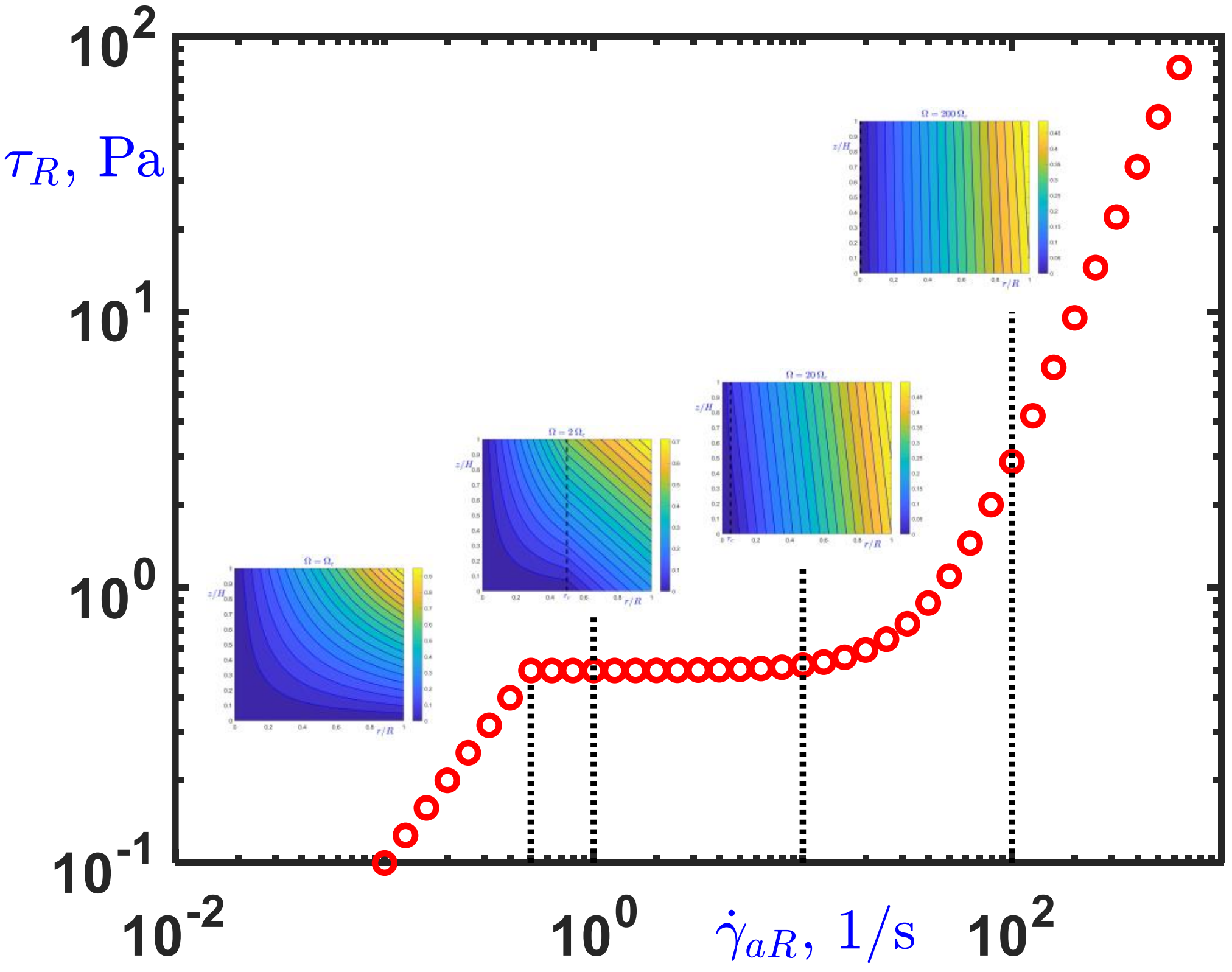

**(a)**

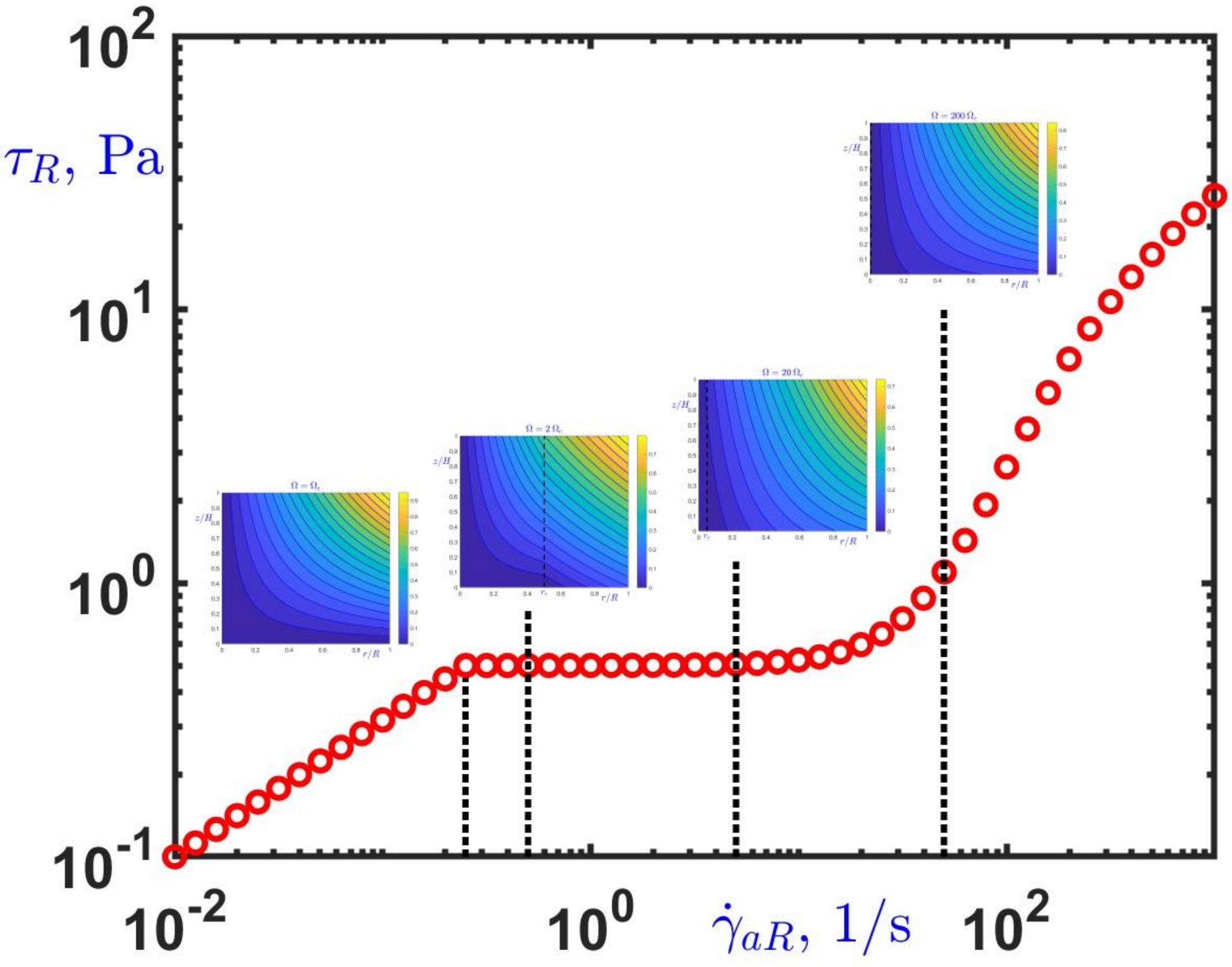

**(b)**

**Figure 7**. Velocity contours for (a) a Newtonian fluid and (b) a power-law fluid with $n = 0.5$ at different apparent shear rates when wall slip with non-zero slip yield stress occurs. The material parameters and the contour plots are those provided in Figs. 5 and 6, respectively.

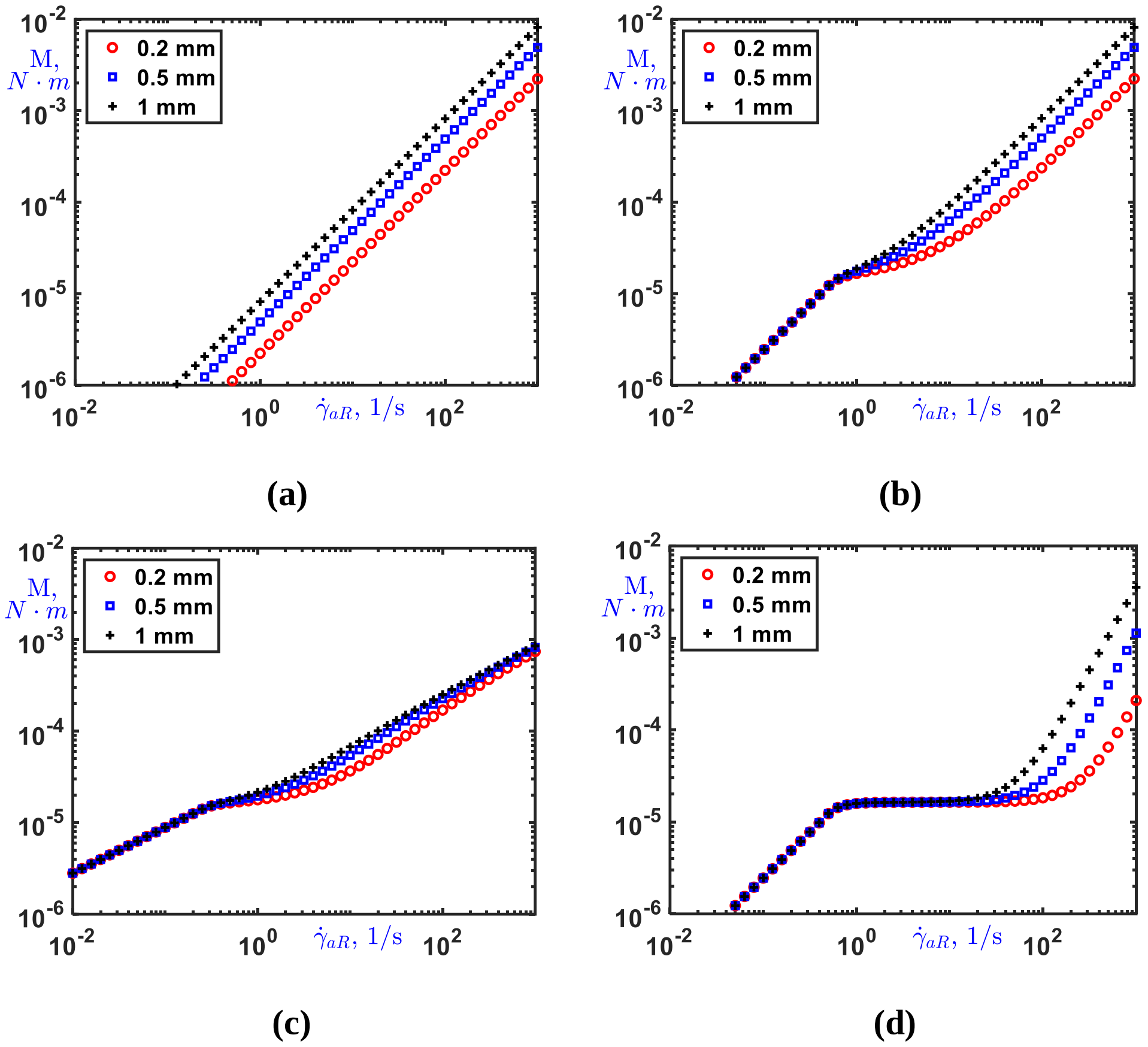

**Figure 8**. Torque versus the apparent rim shear rate in the case of power-law fluids obtained with three gap heights, $H = 0.2$ mm (o), 0.5 mm (□), and 1 mm (∗), $R = 25$ mm, $k = 1$ Pa s$^n$ and $\beta$=1000 Pa s$^m$ / m$^m$: (a) $n$=1, $\tau_c$=0, $m$=1 (Newtonian fluid, zero slip yield stress); (b) $n$=1, $\tau_c$=0.5 Pa, $m$=1 (Newtonian fluid, non-zero slip yield stress); (c) $n$=0.5, $\tau_c$=0.5 Pa, $m$=1 (power-law fluid, non-zero slip yield stress); (d) $n$=1, $\tau_c$=0.5 Pa, $m$=2 (Newtonian fluid, non-zero slip yield stress).

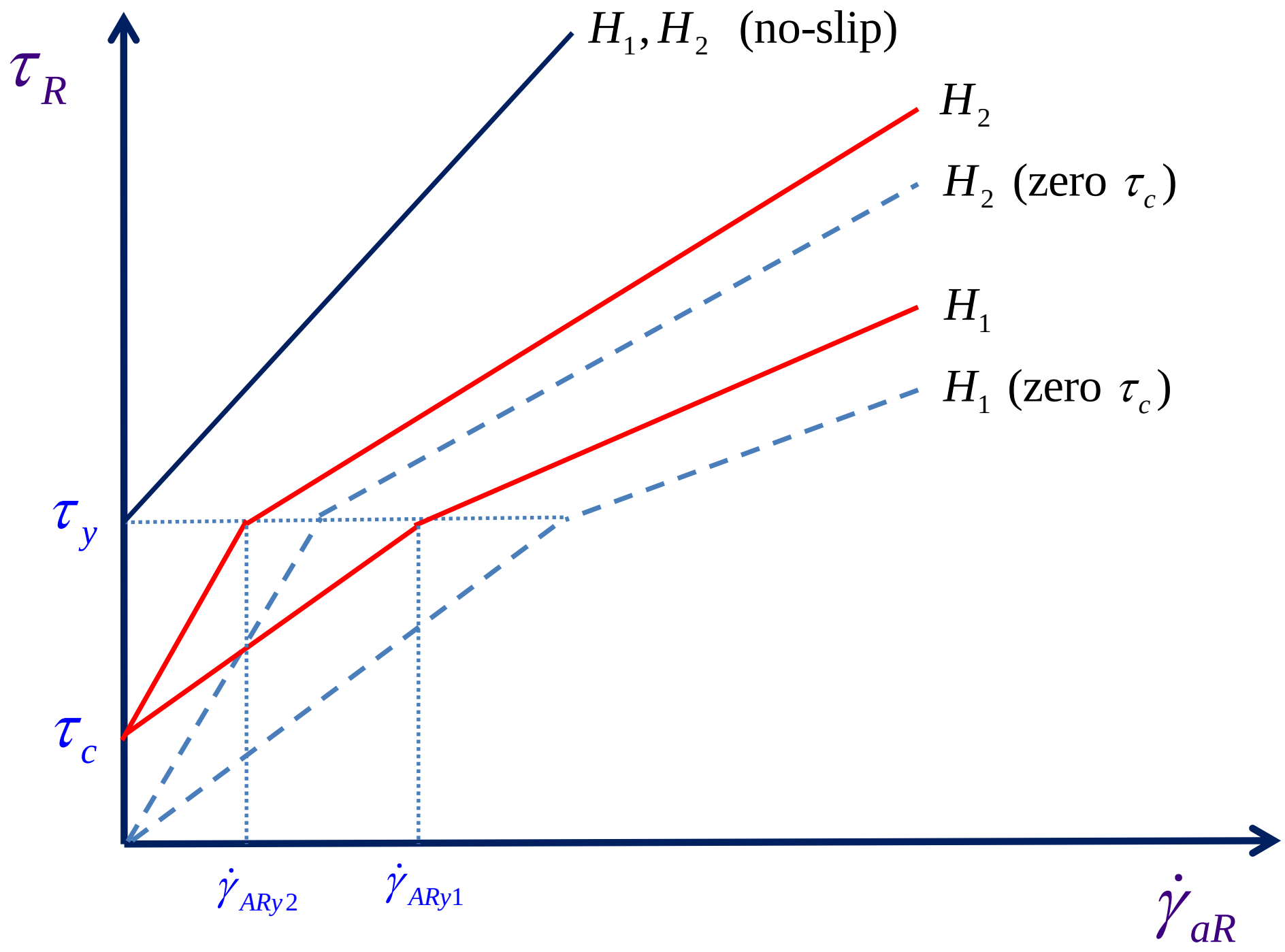

**Figure 9**. Sketch of the gap-size effect ($H_2 > H_1$) on the apparent flow curve of a viscoplastic fluid in the presence of wall slip with non-zero slip yield stress. The dashed lines show the apparent flow curves in the case of zero slip yield stress.

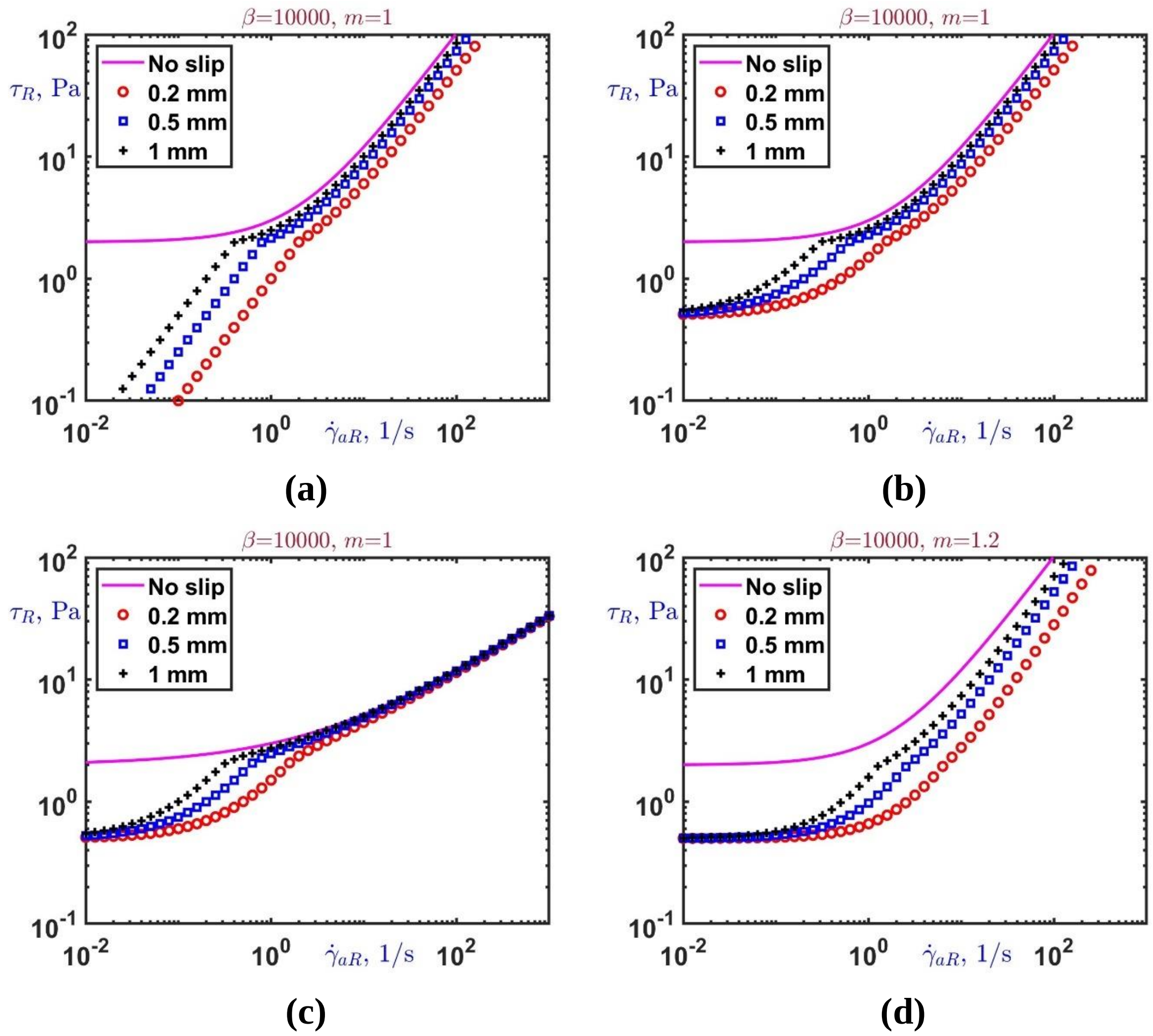

**Figure 10**. Representative apparent flow curves of Bingham and Herschel-Bulkley fluids obtained with three gap heights, $H$ =0.2 mm (o), 0.5 mm (□), and 1 mm (∗), $R = 25$ mm, $\tau_y = 2$ Pa, $k = 1$ Pa s$^n$ and $\beta$=10000 Pa s$^m$ / m$^m$: (a) $n = 1, \ \tau_c = 0, \ \ m = 1$ (Bingham fluid, zero slip yield stress); (b) $n = 1, \ \tau_c = 0.5$ Pa, $\ m = 1$ (Bingham fluid, non-zero slip yield stress); (c) $n = 0.5, \ \tau_c = 0.5$ Pa, $\ m = 1$ (Herschel-Bulkley fluid, non-zero slip yield stress); (d) $n = 1, \ \tau_c = 0.5$ Pa, $\ m = 1.2$ (Bingham fluid, non-zero slip yield stress). The solid line is the flow curve in the case of no-slip.

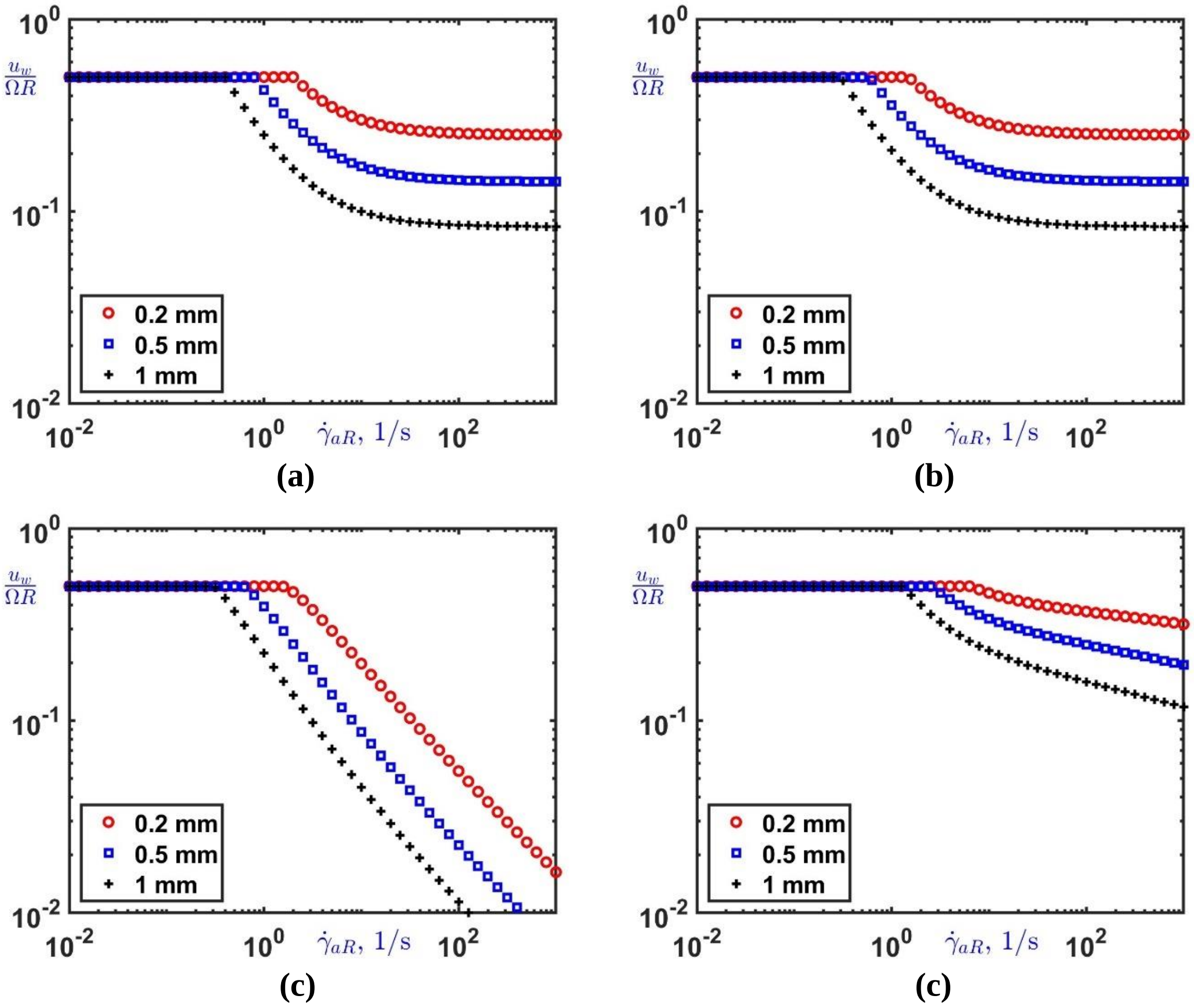

**Figure 11**. Dimensionless rim slip velocities of Bingham and Herschel-Bulkley fluids obtained with three gap heights $(H = 0.2,\ 0.5 \text{ and } 1 \text{ mm})$, $R = 25$ mm, $\tau_y = 2$ Pa, $k = 1$ Pa s$^n$ and $\beta$=10000 Pa s$^m$ / m$^m$ : (a) $n = 1,\ \tau_c = 0,\ \ m = 1$ (Bingham fluid, zero slip yield stress); (b) $n = 1,\ \tau_c = 0.5$ Pa, $m = 1$ (Bingham fluid, non-zero slip yield stress); (c) $n = 0.5,\ \tau_c = 0.5$ Pa, $m = 1$ (Herschel-Bulkley fluid, non-zero slip yield stress); (d) $n = 1,\ \tau_c = 0.5$ Pa, $m = 1.2$ (Bingham fluid, non-zero slip yield stress).

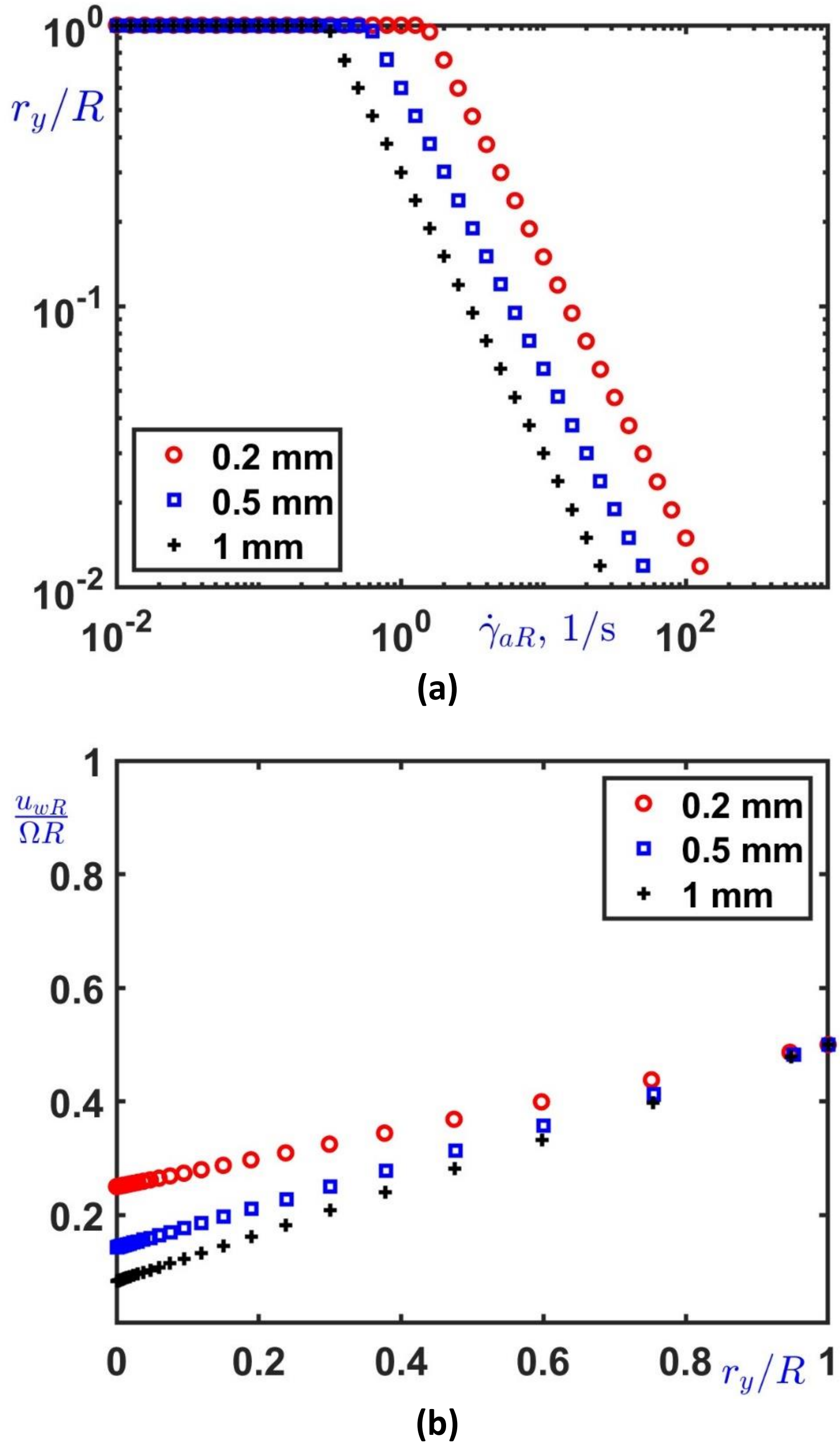

**Figure 12**. (a) Gap size effect on the yield radius $r_y$; (b) Rim slip velocity as a function of the yield radius $r_y$ for different gap sizes; $R = 25$ mm, $\tau_y = 2$ Pa, $k = 1$ Pa s$^n$, $n = 1$, $\tau_c = 0.5$ Pa, $\beta$=10000 Pa s$^m$ / m$^m$, and $m = 1$. The flow curves and the slip velocities for this case are those of Figs. 9b and 10b, respectively.

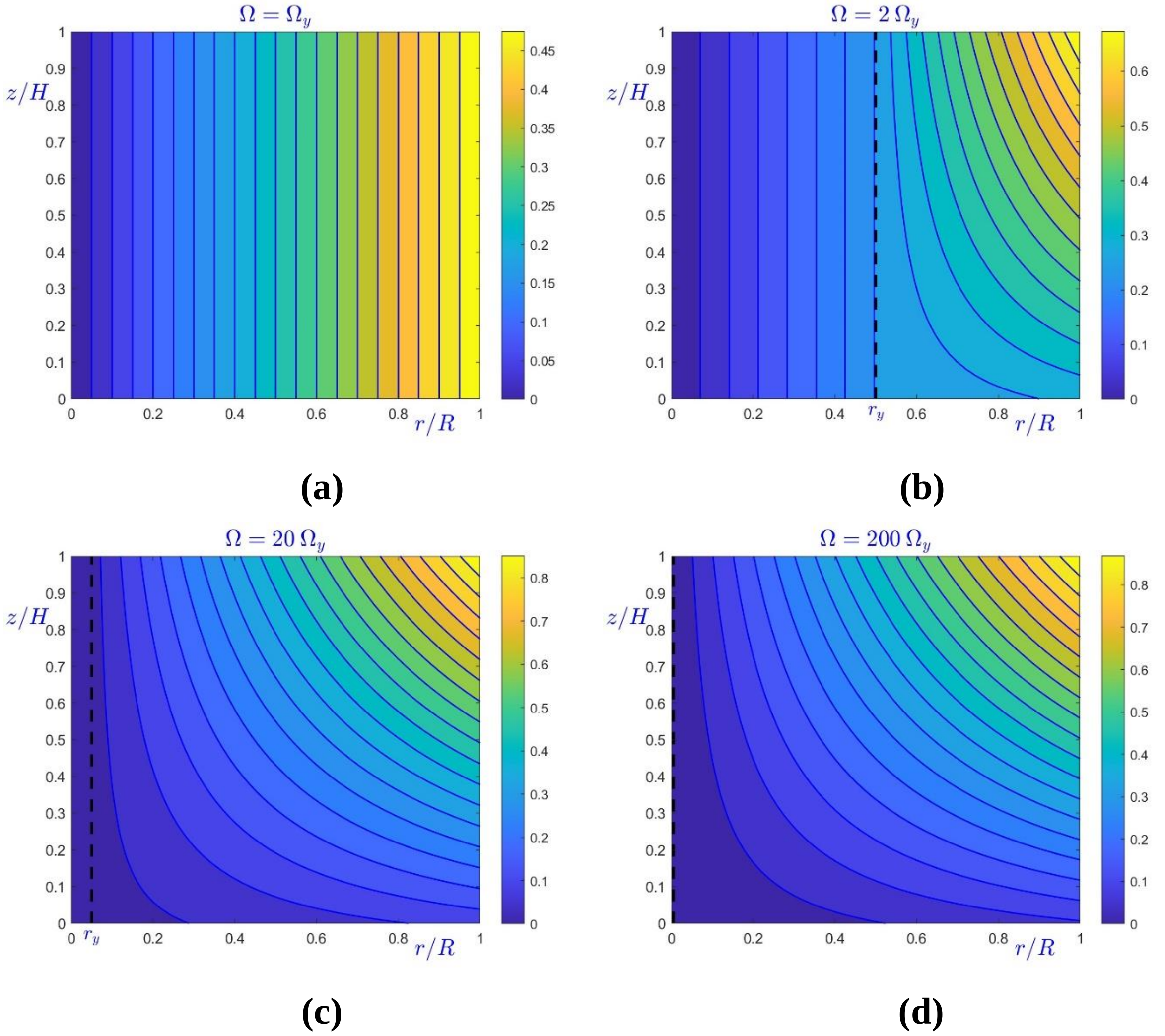

**Figure 13**. Contours of the dimensionless velocity $u_\theta / (\Omega R)$ for various angular velocities in the case of Bingham flow with non-zero slip yield stress, i.e., $R = 25\ \mathrm{mm}$, $H = 1\ \mathrm{mm}$, $k = 1\ \mathrm{Pa\ s}$, $\tau_y = 2\ \mathrm{Pa}$, $\tau_c = 0.5\ \mathrm{Pa}$, $\beta$=10000 Pa s / m, and $m = 1$: (a) $\Omega = \Omega_y$; (b) $\Omega = 2\Omega_y$; (c) $\Omega = 20\Omega_y$; (d) $\Omega = 200\Omega_y$. The critical angular velocity and apparent shear rate are $\Omega_y = 0.012\ \mathrm{s}^{-1}$ and $\dot{\gamma}_{aRy} = 0.3\ \mathrm{s}^{-1}$. The dashed line denotes the yield surface.

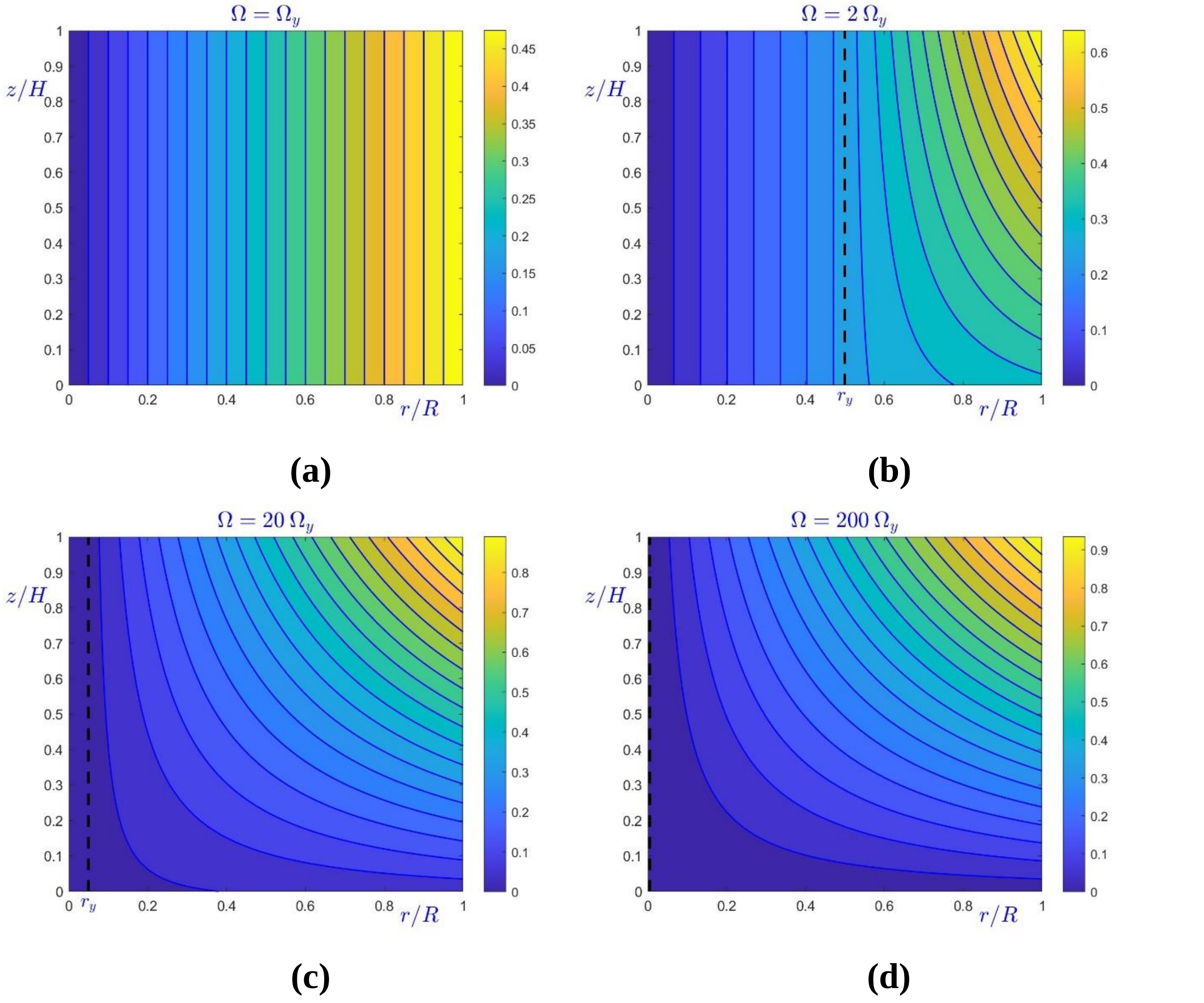

**Figure 14**. Contours of the dimensionless velocity $u_\theta / (\Omega R)$ for various angular velocities in the case of Herschel-Bulkley flow with non-zero slip yield stress, i.e., $R = 25\ \mathrm{mm}$, $H = 1\ \mathrm{mm}$, $n = 0.5$, $k = 1\ \mathrm{Pa\,s}$, $\tau_y = 2\ \mathrm{Pa}$, $\tau_c = 0.5\ \mathrm{Pa}$, $\beta$=10000 $\mathrm{Pa\,s/m}$, and $m = 1$: (a) $\Omega = \Omega_y$; (b) $\Omega = 2\Omega_y$; (c) $\Omega = 20\Omega_y$; (d) $\Omega = 200\Omega_y$. The critical angular velocity and apparent shear rate are $\Omega_y = 0.012\ \mathrm{s}^{-1}$ and $\dot{\gamma}_{aRy} = 0.3\ \mathrm{s}^{-1}$. The dashed line denotes the yield surface.

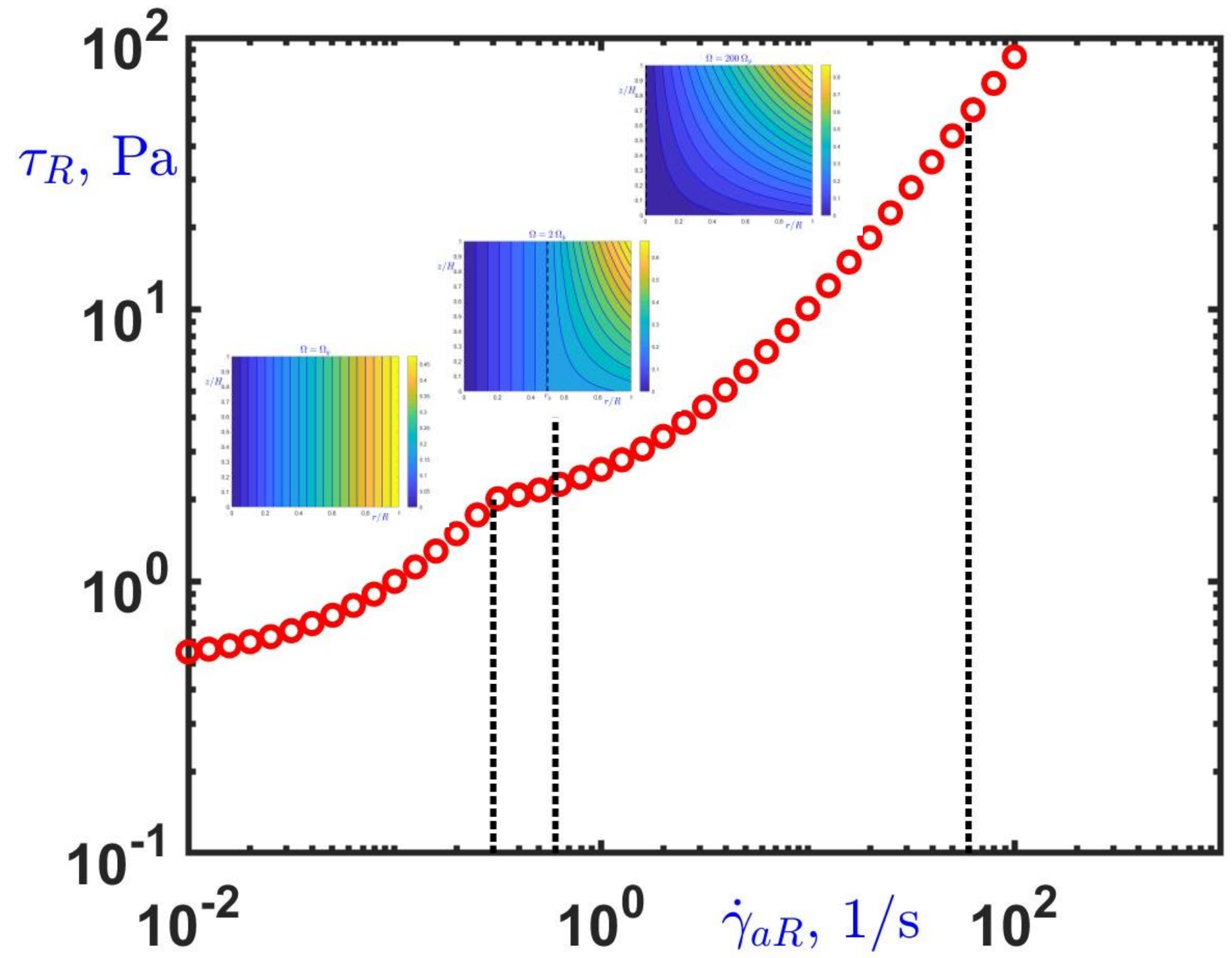

**(a)**

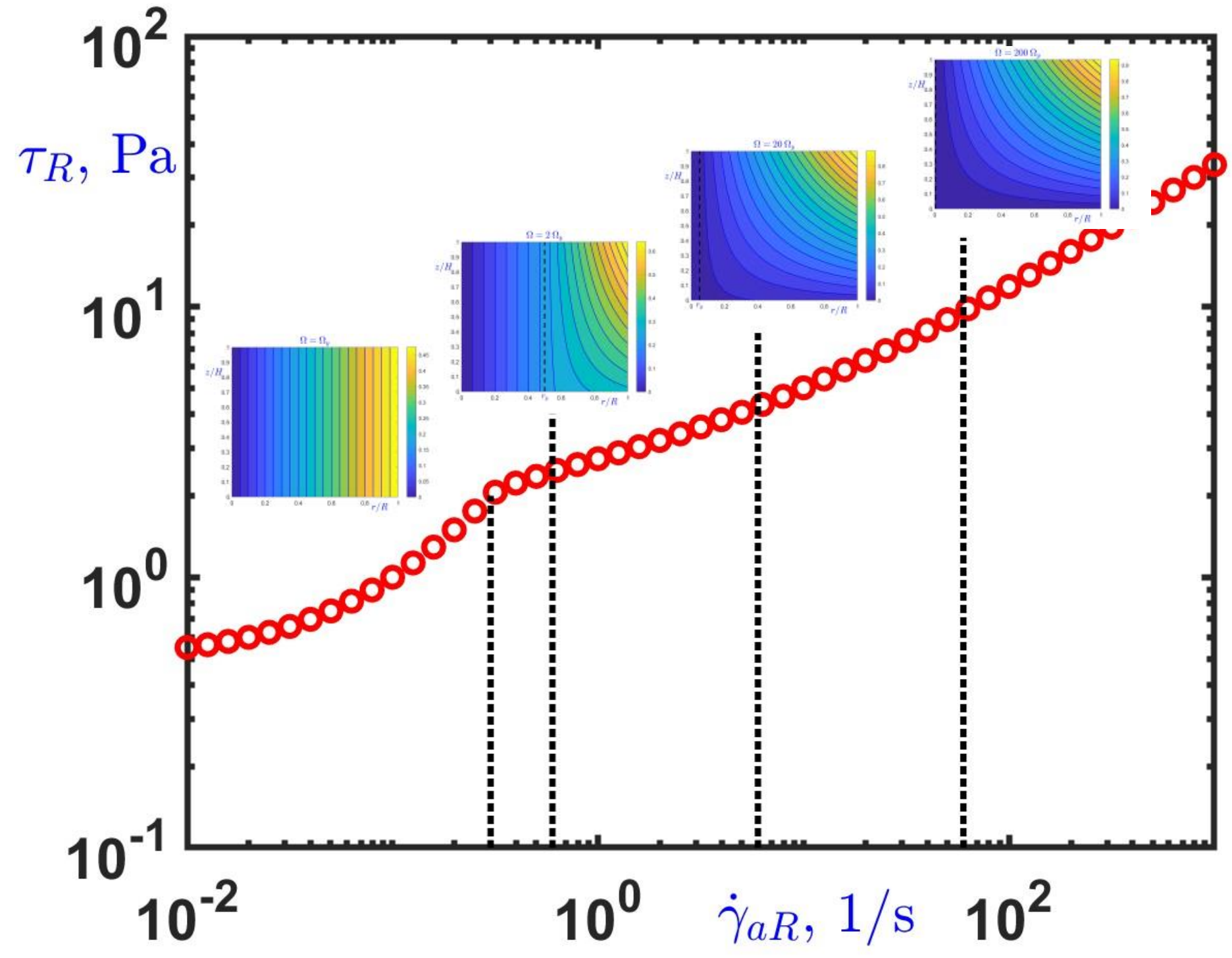

**(b)**

**Figure 15**. Velocity contours for (a) a Bingham fluid and (b) a Herschel-Bulkley fluid with $n = 0.5$ at different apparent shear rates when wall slip with non-zero slip yield stress occurs. Solid body rotation is observed in the first flow regime. The first plateau corresponds to the slip yield stress and the second one to the yield stress.

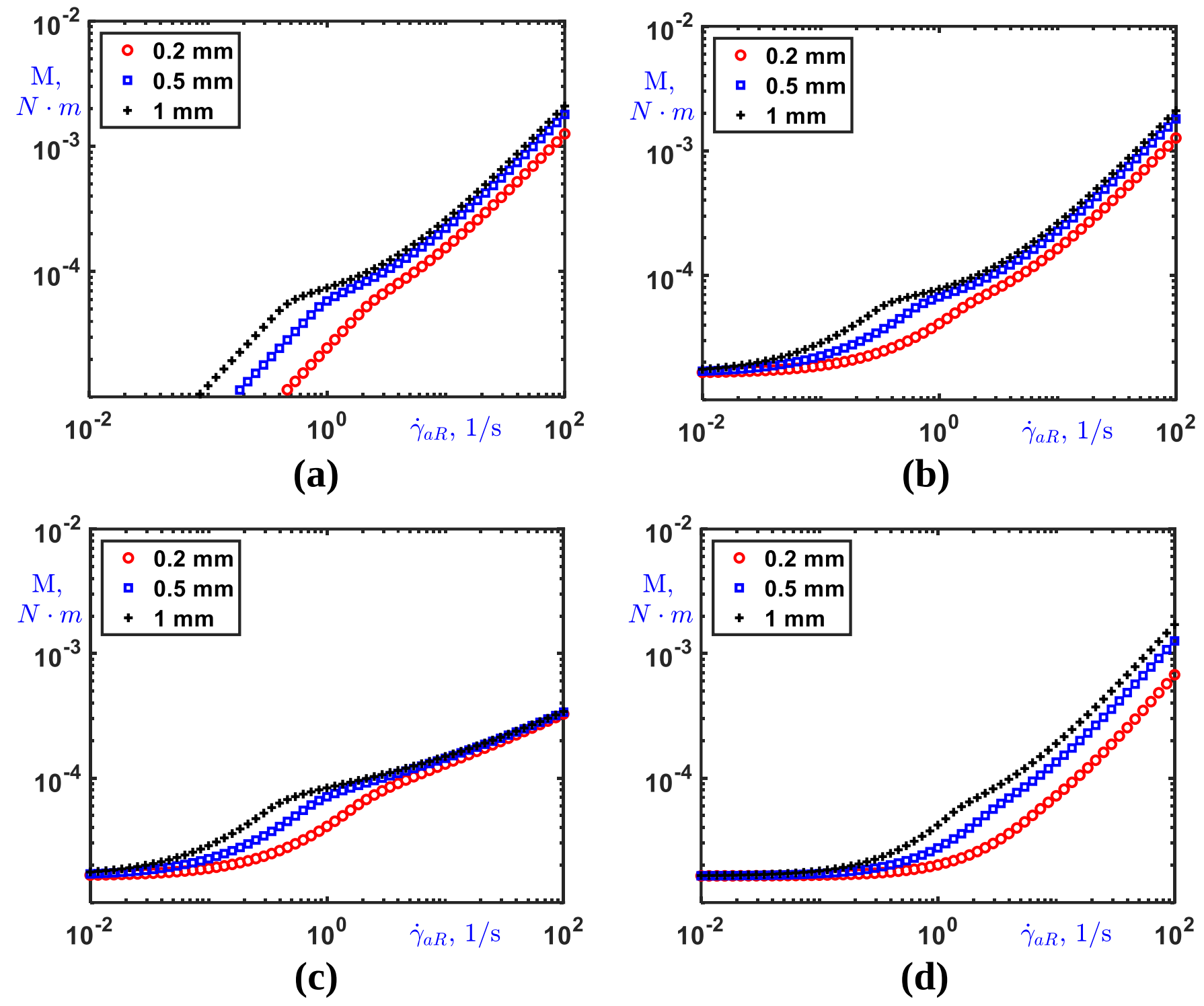

**Figure 16**. Torque versus the apparent rim shear rate in the case of Herschel-Bulkley fluids for three gap sizes, $H = 0.2$ mm (o), 0.5 mm (□), and 1 mm (∗), $R$ =25mm, $\tau_y$ =2Pa, $k = 1$ Pa s$^n$ and $\beta = 10000$ Pa s$^m$/$m^m$: (a) $n$=1, $\tau_c$=0, $m$ =1 (Bingham fluid, zero slip yield stress); (b) $n$=1, $\tau_c$=0.5 Pa, $m$ =1 (Bingham fluid, non-zero slip yield stress); (c) $n$=0.5, $\tau_c$=0.5 Pa, $m$ =1 (Herschel-Bulkley fluid, non-zero slip yield stress); (d) $n$=1, $\tau_c$=0.5 Pa, $m$ =1.2 (Bingham fluid, non-zero slip yield stress).